\documentclass[journal]{IEEEtran}

\usepackage{cite}
\usepackage{url}

\usepackage{textcomp}
\usepackage{xcolor}
\usepackage{hyperref}

\usepackage{booktabs}	
\usepackage{diagbox}    
\usepackage{multirow}   

\usepackage{verbatim}	

\usepackage{makecell}   

\usepackage{amsmath,amssymb,amsfonts,graphicx}
\usepackage{epstopdf}
\usepackage{stfloats} 

\usepackage{algpseudocode}
\usepackage[ruled,linesnumbered]{algorithm2e}
\usepackage{enumitem}
\usepackage{tabularx}

\newtheorem{definition}{Definition}[section]

\newtheorem{upper bound}{Upper bound}

\def\BibTeX{{\rm B\kern-.05em{\sc i\kern-.025em b}\kern-.08em
		T\kern-.1667em\lower.7ex\hbox{E}\kern-.125emX}}

\usepackage{amssymb}
\usepackage{hyperref}
\usepackage{float}

\begin{document}
	
\title{Geospatial Big Data: Survey and Challenges}	
	
\author{Jiayang Wu, Wensheng Gan*, Han-Chieh Chao, Philip S. Yu,~\IEEEmembership{Life Fellow,~IEEE} 
		
\thanks{This research was supported in part by the National Natural Science Foundation of China (No. 62272196), Natural Science Foundation of Guangdong Province (No. 2022A1515011861), and the Young Scholar Program of Pazhou Lab (No. PZL2021KF0023). (Corresponding author: Wensheng Gan)}

\thanks{Jiayang Wu and Wensheng Gan are with the College of Cyber Security, Jinan University, Guangzhou 510632, China; and with Pazhou Lab, Guangzhou 510330, China. (E-mail: csjywu1@gmail.com, wsgan001@gmail.com)}	

\thanks{Han-Chieh Chao is with the Department of Electrical Engineering, National Dong Hwa University, Hualien, Taiwan, R.O.C (E-mail: hcc@ndhu.edu.tw)} 

\thanks{Philip S. Yu is with the University of Illinois Chicago, Chicago, USA. (E-mail: psyu@uic.edu)}  %
		
}

\maketitle

\begin{abstract}	

In recent years, geospatial big data (GBD) has obtained attention across various disciplines, categorized into big earth observation data and big human behavior data. Identifying geospatial patterns from GBD has been a vital research focus in the fields of urban management and environmental sustainability. This paper reviews the evolution of GBD mining and its integration with advanced artificial intelligence (AI) techniques. GBD  consists of data generated by satellites, sensors, mobile devices, and geographical information systems, and we categorize geospatial data based on different perspectives. We outline the process of  GBD  mining and demonstrate how it can be incorporated into a unified framework. Additionally, we explore new technologies like large language models (LLM), the Metaverse, and knowledge graphs, and how they could make GBD even more useful. We also share examples of GBD helping with city management and protecting the environment. Finally, we discuss the real challenges that come up when working with GBD, such as issues with data retrieval and security. Our goal is to give readers a clear view of where GBD mining stands today and where it might go next.
\end{abstract}
	

\IEEEpeerreviewmaketitle

\section{Introduction} \label{sec:introduction}

In the digital era \cite{gan2023web,wan2023web3}, integrating cutting-edge technologies has resulted in the development of  GBD mining. GBD refers to the massive and complex data generated by various sources such as satellites, sensors, mobile devices, social media, and geographical information systems \cite{lee2015geospatial}. These data consist of geographic, temporal, and attribute information. The impact of GBD is reshaping how we perceive, analyze, and interact with spatial information on an unprecedented scale. The significance of GBD mining cannot be underestimated. As the volume and variety of geospatial data continue to rise, extracting meaningful knowledge from this data becomes a huge challenge \cite{lee2011recent}. GBD mining transcends traditional data analysis methodologies, offering a gateway to unlock hidden patterns in geospatial information. It arms decision-makers, researchers, and policymakers with essential tools. These tools help them grasp urban transformations, environmental shifts, disaster occurrences, and socio-economic changes.

Artificial intelligence (AI) \cite{brynjolfsson2017artificial} plays a crucial role in GBD mining by providing the cognitive capacity to tackle the complexity of this data. AI techniques, including machine learning, deep learning, and natural language processing (NLP), enable the automated extraction from GBD. AI-driven GBD mining improves spatial data analysis \cite{shekhar2010spatial}, predictive modeling \cite{wang2020deep}, pattern recognition \cite{mennis2009spatial}, anomaly detection \cite{lu2003algorithms}, and advanced visualization \cite{keim2004visual}. By leveraging AI, GBD mining extends beyond traditional data processing, transforming raw geospatial data into practical knowledge \cite{zhao2021deep}.

To integrate these advanced AI techniques for effective mining of various types of GBD, it is important to utilize a unified framework that includes every stage of data handling, from preprocessing, storage, and retrieval to analysis, prediction, and visualization. This integration requires each step to be linked, ensuring the harmonious operation of AI technologies within a singular system. The potential combination of emerging technologies such as LLM \cite{wu2023multimodal}, knowledge graphs, and the Metaverse could introduce revolutionary improvements to GBD mining. LLM could enhance the understanding and generation \cite{han2023chartllama}, while knowledge graphs might provide a structured representation of geospatial entities and their relationships \cite{mai2022symbolic}, facilitating complex queries and analyses. The Metaverse could offer immersive, three-dimensional visualization capabilities, interpreting geospatial data more intuitive \cite{zauskova2022visual}. In terms of application scenarios, urban management \cite{jiang2010geospatial} and environmental sustainability \cite{rai2022geospatial} stand out as areas that could greatly benefit from the advancements in GBD mining. In urban management, the GBD mining could improve city planning \cite{shen2012geospatial}, traffic management \cite{jiang2018geospatial}, and emergency response \cite{brunner2009distributed}. For environmental sustainability, this approach could aid in monitoring climate change \cite{avtar2020utilizing}, managing natural resources \cite{lacroix2019mapx}, and predicting environmental disasters \cite{hong2007experimental}.

Building upon the unified framework for GBD mining and the integration of advanced technologies, real-world applications present additional challenges that must be addressed, notably the increasing scale of data and the critical importance of data security. The volume of digital data is expanding at a remarkable pace, with statistics indicating that the digital domain doubles in size every two years, projected to reach 175 zettabytes by 2025\footnote{https://spectrum.ieee.org/tape-storage-sustainable-option}. This exponential growth requires innovative solutions for data storage and retrieval to manage and process GBD efficiently. In response to this challenge, leveraging existing storage technologies and indexing techniques could be important, such as the design of vectorized databases and the construction of spatial indexes. On the security front, the increase in data scale brings about risks. It is essential to incorporate security mechanisms to safeguard GBD. Techniques such as federated learning and anomaly detection may offer promising paths for enhancing data security. Federated learning, by allowing data analysis models to be trained on decentralized devices without needing to exchange raw data, can mitigate the risks of data centralization. Meanwhile, anomaly detection algorithms can identify unusual patterns or potential threats within the data, allowing for the detection of malicious data.

\textbf{Research gap}: In the existing literature on GBD, critical insights emerge from studies such as Lee \textit{et al.} analysis of the challenges and opportunities in GBD \cite{lee2015geospatial} and Hu \textit{et al.} \cite{hu2016resource} design of a big data architecture for the management of remote sensing. These contributions, along with NoSQL database storage solutions \cite{li2017geo}, data processing technologies \cite{hu2016resource}, and effective data management and analysis \cite{al2021review}. While existing literature reviews have significantly contributed to the understanding and application of GBD mining, these studies often emphasize specific aspects of the field rather than providing a holistic view. This survey aims to bridge this gap by emphasizing a holistic view of GBD mining, focusing on the combination of different stages in the mining processing and how these can be integrated into a unified framework. What's more, this paper also gives insight into the incorporation of novel technologies such as LLM, Metaverse, and knowledge graphs, and how their application can significantly enhance the capabilities in handling GBD. In addition, through detailed examples, this survey shows the application scenarios of GBD in urban management and environmental sustainability. Finally, this paper lists the practical challenges encountered in the applications of GBD mining and offers insights into issues related to data retrieval and security.

\textbf{Contributions}: The major contributions of this survey paper are shown as follows:
	
\begin{itemize}
    \item We define key definitions related to GBD. It consists of the description of GBD including geospatial data, its difference from geographic data, geospatial analysis, and GeoAI.  It also includes the introduction of important technologies like remote sensing, GPS, GIS, data processing, data visualization, databases, and mobile location. These definitions lay the foundation for understanding GBD and its applications (Section \ref{sec:concept}).

    \item We categorize geospatial data into three main aspects, including data types, data formats, and the sources of geospatial data. This classification provides a comprehensive overview of the diverse geospatial data (Section \ref{sec:data}).
    
    \item  We offer a comprehensive description of the GBD mining process (Section \ref{sec:process}). This includes multiple steps ranging from data preprocessing, storage, retrieval, analysis, and prediction to visualization. Furthermore, we present an integration of these phases into a unified framework.

    \item We explored the use of LLM (Section \ref{sec:LLM}) and emphasized the importance of GBD for urban management and environmental sustainability (Section \ref{sec:cities} and Section \ref{sec:enviromental}).
   
    \item We discuss challenges in GBD mining (Section \ref{sec:challenges}), including data retrieval optimization and privacy concerns. We also highlight several future directions (Section \ref{sec:future}), including GBD database design,  heterogeneous data management, integration with knowledge graphs, visualization, and privacy protection.    
\end{itemize}

\begin{figure}[ht]
    \centering
    \includegraphics[scale=0.41]{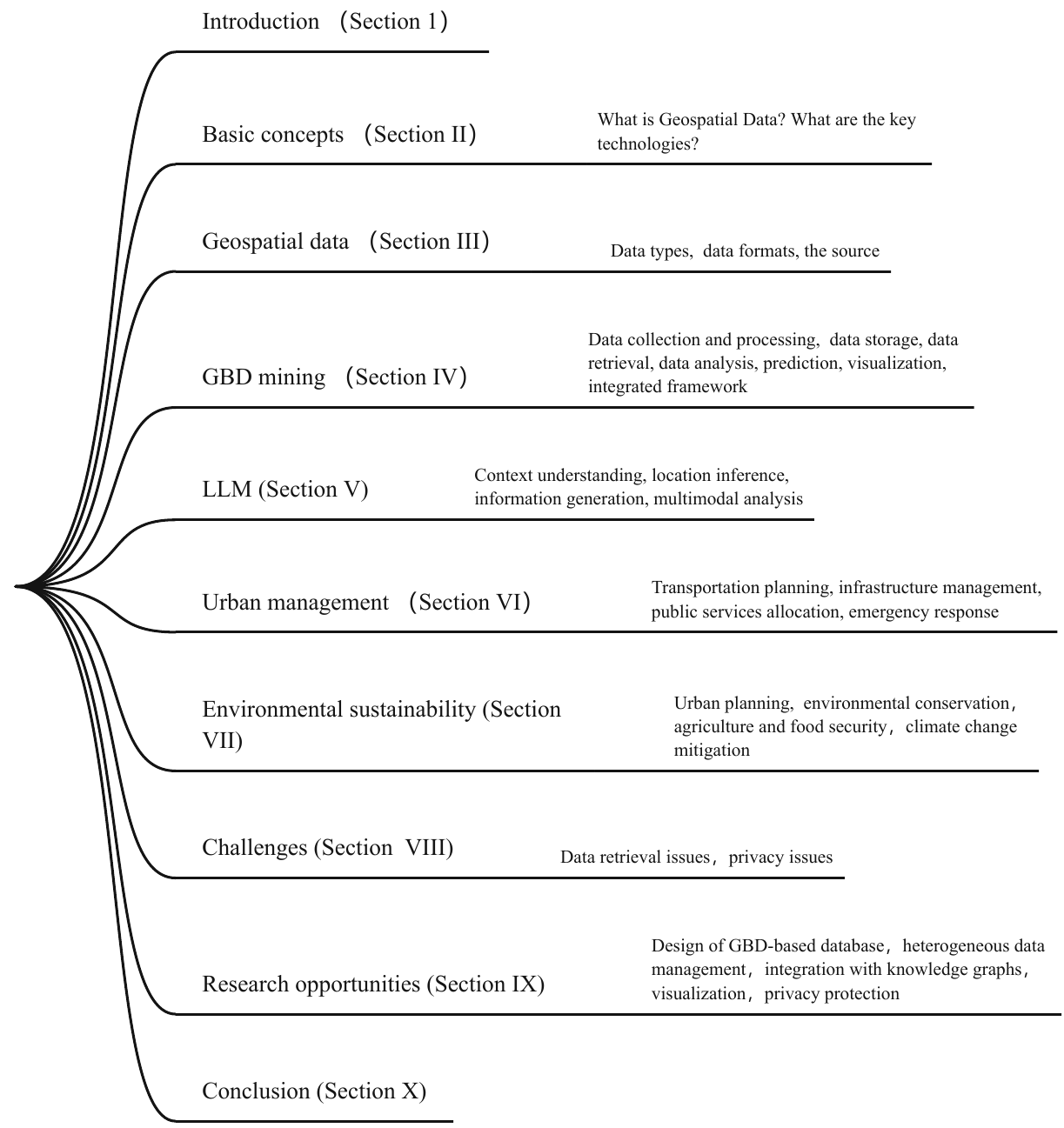}
    \caption{The outline of our overview.}
    \label{fig:outline}
\end{figure}

\textbf{Organization}:  The organization of this review is shown in Fig. \ref{fig:outline}. In Section \ref{sec:concept}, we present the basic concepts and techniques related to GBD. We discuss the type of data used in GBD in Section \ref{sec:data}. Furthermore, we explore advanced technologies for GBD mining in Section \ref{sec:process}. We highlight the significance of technology advancement on data capture via LLM in Section \ref{sec:LLM}. In addition, we list the examples of urban management in Section \ref{sec:cities} and environmental sustainability in Section \ref{sec:enviromental}. In Section \ref{sec:challenges} and Section \ref{sec:future}, we propose some challenges and research opportunities. Finally, we conclude this survey in Section \ref{sec:conclusion}. 

\section{Basic Concepts}  \label{sec:concept}
	
The frequently used symbols in this paper are summarized in Table \ref{tab:summary_symbols}, and the relevant definitions are presented below.

\begin{table}[ht]
    \caption{Summary of Symbols and Corresponding Explanations}
    \centering
    \label{tab:summary_symbols}
    \begin{tabular}{lp{6.5cm}}
        \toprule   
        \textbf{Symbol} & \textbf{Definition} \\		
		\midrule  
        GBD &  Geospatial big data. \\
        GPS &  Global position system. \\
        GIS &  Geographic information science. \\
        AI & Artificial intelligence. \\ 
        LLM & Large-scale language model. \\
        GeoAI &  Geospatial artificial intelligence.\\
        \bottomrule
    \end{tabular}
\end{table}
	
\subsection{What is Geospatial Data?}

\begin{definition}[Geospatial data]
  \rm  Geospatial data possesses the information that describes objects, events, or features located on or in close proximity to the Earth's surface \cite{bertino2008security}. It merges location specifics, represented as Earth coordinates, with attribute details characterizing associated objects, along with temporal data indicating the duration of their existence \cite{triglav2011spatio}. Locations can be either static, such as equipment or earthquake events, or dynamic, including moving entities like vehicles, pedestrians, or the spread of infectious diseases. Geospatial data comprises datasets gathered in various formats from multiple sources, including satellite imagery, weather records, cell phone data, mapped visuals, and social media content. Its full potential emerges when it can be discovered and analyzed.
\end{definition}
  
\begin{definition}[Geospatial data vs. geographic data]
  \rm  Geospatial data consists of any data type that describes phenomena related to the geospatial dimension. In contrast, geographic data specifically refers to geospatial data with a dimension linked to its position either at a particular moment or on the Earth's surface \cite{anselin2009geoda}.
\end{definition}

\begin{definition}[Geospatial analysis]
  \rm  Geospatial analysis enhances traditional data by incorporating time and location information, facilitating the creation of informative data visualizations. These visual representations consist of maps, charts, statistics, and graphs, effectively conveying historical and temporary trends \cite{murayama2012progress,de2007geospatial}. The accompanying context enriches event descriptions, often revealing insights that might have been obscured in tables of data. Clear and familiar visual patterns and images speed up understanding, making predictions faster and more accurate.
\end{definition}

\begin{definition}[Geospatial technology]
  \rm  Geospatial technology consists of a wide range of technologies designed for the acquisition, storage, and management of geographic information \cite{goodchild2009geographic}. Geospatial technology interfaces with several interconnected technologies, including GIS, GPS, and remote sensing. In addition, it includes satellite technology, which plays an important role in geographic mapping and Earth analysis. Moreover, GIS, in particular, focuses on the visual representation and spatial mapping of data. For instance, the practical use of GIS is clear when we look at a hurricane map that shows both location and time. 
\end{definition}
 
\begin{definition}[Geospatial intelligence]
  \rm  GeoAI refers to the utilization of high-performance computing tools, the application of AI technologies, as well as advanced data mining methods \cite{dold2017future}. GeoAI is new and important and incorporates recent developments in geospatial science. This field aims to extract meaningful insights, patterns, and knowledge from vast geospatial data, enabling decision-making across various domains such as urban planning, environmental monitoring, and disaster management.
\end{definition}

\subsection{What are the key technologies?}
Geospatial data involves various technologies and methods that are used for the collection, processing, analysis, and visualization of geospatial data. Here are some key technologies involved:

\begin{definition}[Remote sensing]
  \rm  Remote sensing involves the capture of Earth's surface imagery and data utilizing remote sensors like satellites and aircraft \cite{aggarwal2004principles}. It includes different ways of collecting information, like using cameras, radar, and infrared technology to gather various types of geospatial data. This data can point to things like what's on the ground, the weather, plants, and many other details.
\end{definition}

\begin{definition}[GPS]
  \rm  GPS technology plays a crucial role in determining and measuring longitude, latitude, and elevation data at any point on Earth's surface \cite{kursinski1997observing}. It has widespread applications in geospatial data collection, vehicle tracking, and navigation, facilitating precise positioning and location-based services.
\end{definition}

\begin{definition}[GIS]
  \rm GIS technology is employed for storing, managing, analyzing, and visualizing geospatial data \cite{lacroix2019mapx}. It is used for map-making, geospatial data storage and querying, and geospatial analysis, serving as a core tool for processing geospatial data.
\end{definition}

\begin{definition}[Data processing]
  \rm Data processing is the systematic and organized manipulation of geospatial data to extract meaningful information or transform it into a more usable format \cite{monteiro2018urban}. It involves a series of operations such as data cleaning, aggregation, analysis, and interpretation to support decision-making and problem-solving.
\end{definition}

\begin{definition}[Data visualization]
  \rm Data visualization refers to the representation of geospatial data in a visual format, often using charts, graphs, maps, or interactive visualizations \cite{dodge2008power}. The goal of data visualization is to make complex geospatial datasets more understandable by revealing trends, and relationships that might be less apparent in raw data. It is important in conveying geospatial information to both technical and non-technical audiences.
\end{definition}

\begin{definition}[Geospatial database]
  \rm A geospatial database is a technology that stores and manages large-scale geospatial data \cite{ziliaskopoulos2000internet}. These databases are specifically designed to support geospatial indexing, geospatial querying, and geospatial analysis. They enable the storage of geographic and attribute data, making it possible to perform spatial queries and geospatial analyses for various applications, including GIS, location-based services, and spatial data science.
\end{definition}

\begin{definition}[Mobile location]
  \rm Mobile location is used to obtain precise location information from mobile devices such as smartphones, tablets, and GPS receivers \cite{tsou2004integrated}. It relies on various sources of information, like satellite signals and sensors inside devices, to figure out and follow the exact geographic position of mobile devices as they move. It includes their latitude (how far north or south they are), longitude (how far east or west they are), and often their elevation (how high above sea level they are). This technology enables a variety of location-based features and applications.
\end{definition}

\section{Geospatial Data} \label{sec:data}

\begin{figure}[htbp]
	\centering
	\includegraphics[width=0.48\textwidth]{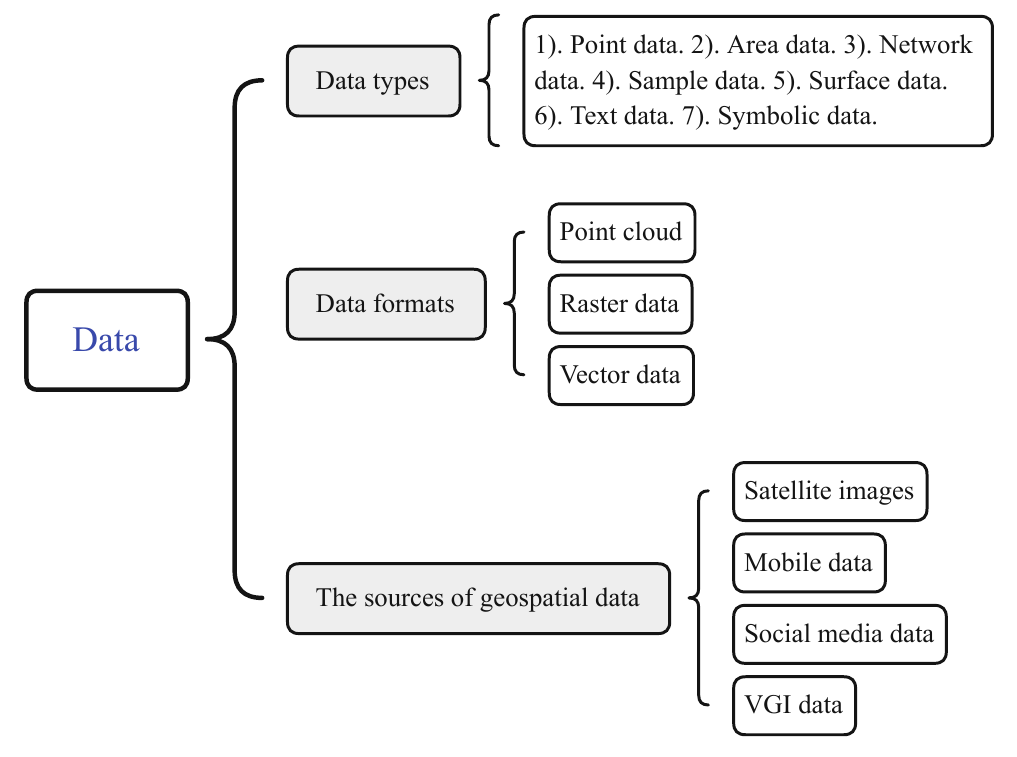} 
	\caption{The types of data.}
	\label{data} 
\end{figure}

Geospatial data can be classified in various ways, including data types, data formats, and the sources of geospatial data, as shown in Fig. \ref{data}.

\subsection{Data types} 
According to its structural characteristics \cite{burrough2015principles}, geospatial data representing geographic phenomena can be categorized into various types, including point, area, network, sample, surface, text, and symbolic data, which are represented using coordinates or geographic methods.

\subsection{Data formats}

\begin{figure}[htbp]
    \centering
    \includegraphics[width=0.45\textwidth]{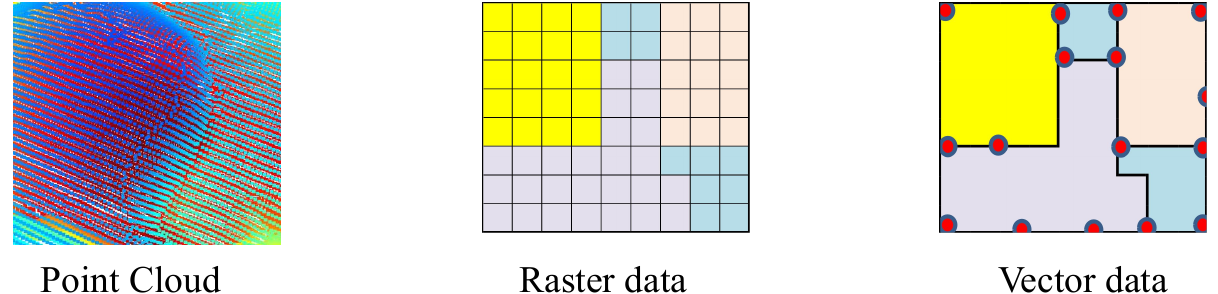} 
    \caption{Some examples of the geospatial data.}
    \label{data_format} 
\end{figure}

Geospatial data can be classified into three formats \cite{cura2017scalable}: point cloud, raster data, and vector data, as shown in Fig. \ref{data_format}.

\textbf{Point cloud.} Point cloud data comprises a collection of spatially co-located points, often acquired through 3D scanning or remote sensing techniques \cite{han2017review}. These points collectively form a detailed 3D model of the scanned object or environment.

\textbf{Raster data.} It involves pixelated or gridded cells, identified by rows and columns. Images generated from raster data \cite{brisaboa2017efficiently}, including photographs and satellite images.

\textbf{Vector data.} It represents geographic features as precise points, lines, or polygons \cite{xavier2016survey}. Vector data is highly precise and versatile, making it suitable for various spatial analyses and map-making tasks.

\subsection{The sources of geospatial data.}

Geospatial data originates from a diverse range of sources, each offering unique insights and capabilities for analysis. These include satellite images, mobile data, social media data, and volunteered geographic information (VGI).

\textbf{Satellite images.} These images provide a comprehensive view, capturing the fine details of our planet's surface \cite{geman1996active}. They are invaluable for a wide range of applications, from environmental monitoring to urban planning.

\textbf{Mobile data.} This type of data is generated by mobile devices and includes information such as call records, text message logs, and internet usage data \cite{sarker2021mobile}. When combined with GPS technology, mobile data can provide insights into human mobility patterns, allowing researchers to study how people move and interact in different environments. 

\textbf{Social media data.} With millions of users worldwide, social media platforms generate vast amounts of data that can be geo-referenced to specific locations \cite{naaman2011geographic}.

\textbf{VGI data.} VGI data is a form of crowd-sourced data where individuals voluntarily contribute geographic information \cite{fonte2017assessing}. It allows users to add, edit, and share geographic data. This results in detailed and up-to-date maps of areas that might be poorly covered by traditional mapping agencies.

\section{Geospatial data mining} \label{sec:process}

\subsection{Data collection and processing}

This subsection shows the steps involved, which include sourcing raw data, preparing it for analysis, and harnessing advanced technologies to ensure efficient processing. 

\textbf{Data sources.} The sources can be quite varied, like satellite images, GPS data, aerial photos, databases, or even information from the Internet \cite{raup2007glims}. Each source has its characteristics, from detailed satellite pictures to real-time GPS location data. Because the quality of these sources can differ, it's important to handle the data carefully.  This means cleaning it up, filtering out what is not needed, making sure all the data is in a standard format, and organizing it, so we can quickly find what we're looking for. This is the first step in making sense of all this geospatial information.

\textbf{Data preprocessing.} After understanding the data sources, we should prepare the data for in-depth analysis. This entails several tasks, such as cleaning up the data to remove errors, filtering out irrelevant information, ensuring that all data follows a common format, and organizing it for retrieval \cite{blazquez2018big}. 

\textbf{Distributed processing.} Apache Spark, a distributed computing framework, enables the efficient handling of large-scale geospatial datasets \cite{lunga2020apache}. It offers the capability to process and analyze geospatial information in parallel across multiple nodes, enhancing speed and scalability. Apache Spark's distributed processing capabilities can significantly accelerate geospatial data analysis.

\textbf{Geospatial data fusion.} It involves combining data from multiple sources to gain a more comprehensive understanding of a geographical phenomenon \cite{zhang2016multisource}. By merging data from various sensors, databases, or satellite sources, analysts can create more accurate geospatial datasets. This fusion process unlocks deeper insights and enables a more holistic approach to geospatial analysis.

\subsection{Data Storage}

Efficiently managing geospatial data is integral to unlocking the potential insights it holds.  This section provides a comprehensive overview of their significance, showing how they serve for managing, querying, and analyzing vast geospatial information. We will first present the basic technologies for database storage. In addition, we will give some examples of data storage for specific types of data.

\textbf{Geospatial indexing.} Geospatial indexing techniques include r-tree \cite{goyal2020grid}, quadtree \cite{mathew2010quad}, and grid index, designed to enhance geospatial queries' efficiency. These methods could help optimize data retrieval. They provide spatial data structures that enable quicker and more accurate retrieval of geospatial information. The choice of indexing technique often depends on the data and the specific requirements of the geospatial application.

\textbf{NoSQL geospatial databases.} NoSQL databases like MongoDB and Cassandra support geospatial data storage \cite{baruffa2019comparison}. They enable spatial indexing and efficient querying of geospatial information. NoSQL geospatial databases have become important tools for organizations across various sectors. They utilize these tools to manage the volume of geospatial data such as transportation, agriculture, urban planning, and environmental monitoring.

\textbf{Image data storage.} Geospatial image data can be efficiently managed through tile-based storage and cloud storage solutions \cite{gorelick2017google}. Efficient management of geospatial image data is crucial for geospatial applications across various industries, including agriculture, environmental monitoring, disaster management, and urban planning. Tile-based storage and cloud storage solutions like AWS \cite{shankar2020serverless}, Google Cloud \cite{gorelick2017google}, and Azure \cite{teng2010medical} provide the necessary infrastructure and tools to handle large-scale geospatial image datasets. 

\textbf{Vector data storage.} Vector geospatial data, which includes points, lines, and polygons \cite{lv2019bim}. Two common technologies for storing vector geospatial data are PostGIS and formats like GeoJSON \cite{agarwal2016performance}. Both PostGIS and GeoJSON offer valuable options for storing vector geospatial data. The choice between them often depends on the specific requirements of the application, the need for complex spatial analysis, and compatibility with existing systems. 

\textbf{Temporal data storage.} Temporal data tracking changes over time is essential. Temporal databases and time-series databases are used for efficient storage and querying of time-stamped geospatial data \cite{george2007spatio}. These databases like InfluxDB \cite{papoutsis2020insar} and TimescaleDB \cite{khelifati2023tsm} are tools for managing and analyzing time-stamped geospatial data. With the ability to efficiently store and query time-stamped geospatial data, organizations can gain deeper insights into how geographic features change over time, enabling better decision-making and resource management.

\textbf{Geospatial data compression algorithms.} When dealing with large-scale geospatial data, geospatial data compression algorithms can reduce storage space and transmission costs. Common geospatial data compression algorithms include grid-based compression methods \cite{walker2001wavelet} and linear compression.

\subsection{Data retrieval}  The retrieval of geospatial data requires the rapid extraction of information from vast repositories. These retrieval techniques are categorized based on data types and specific application contexts. With the help of illustrative examples, we give an overview of several geospatial data retrieval techniques.

\textbf{Attribute-based retrieval.} This retrieval method leverages attribute information to query vector data, such as searching for specific types of locations or finding administrative regions that meet attribute criteria \cite{siddiquie2011image}. For example, one task could be querying all cities with a population exceeding 100,000 from vector data of administrative boundaries.  This approach is fundamental in GIS for filtering data that meet specific, attribute-based criteria.

\textbf{Spatial relationship-based retrieval.} Utilizing geographic spatial relationships, this method queries vector data to find features within a specific area or relative to each other \cite{punitha2005invariant}. For instance, identifying all parks within an administrative region, or finding all bus stops within a specified distance from a given location. This technique is invaluable for understanding spatial arrangements and relationships.

\textbf{Spatio-temporal based retrieval.} For time-series data, spatio-temporal retrieval is crucial. It enables queries based on both spatial and temporal dimensions, such as tracking the temperature trends within a specific region over a chosen time period \cite{chu2021long}. This method supports environmental monitoring insights into changes and patterns over time and space.

\textbf{Feature matching retrieval.}  Applicable to image data, this technique involves computing and indexing image feature vectors, followed by similarity comparison to find matching images. Deep hashing-based image retrieval methods, for example, transform images into binary hash codes \cite{jang2022deep}. It allows for rapid identification of visually similar images by comparing their hamming distances. This method is particularly useful for applications like landscape change detection, land cover classification, and environmental monitoring.

\textbf{GIS-based retrieval.} GIS facilitates the visual and interactive querying of geospatial data \cite{shahabi2010geodec}. Users can perform visual searches by drawing selection boxes, delineating areas, or marking points on a map, swiftly locating the geospatial data of interest. This method enhances user interaction with data.

\subsection{Data analysis}

Geospatial data involves the analysis and processing of geospatial data through feature extraction, clustering, and classification. Various types of algorithms can be employed during this process, including but not limited to the following, also listed in Table \ref{table:1}:

\begin{table*}[ht]
	\caption{The analysis methods for geospatial data.}
	\label{table:1}
	\begin{tabularx}{\textwidth}{|m{3cm}<{\centering}|m{1cm}<{\centering}|m{1cm}<{\centering}|X|}
		\hline
		\textbf{Algorithm} & \textbf{Paper} & \textbf{Year} & \multicolumn{1}{c|}{\textbf{Description}} \\
		\hline
		\textbf{Texture feature extraction} & \cite{lucieer2005multivariate} & 2004 & These algorithms analyze spatial patterns and textures in geospatial images, providing valuable texture-based features for classification. \\
		\hline
		\textbf{Geospatial association rule mining} & \cite{shyu2006knowledge}  & 2006 & These algorithms identify patterns and associations in geospatial data, helping discover interesting spatial relationships. \\
		\hline
		\textbf{Edge detection} & \cite{li2008gis} & 2011 & Edge detection algorithms identify boundaries and edges in geospatial images, aiding in feature extraction. \\
		\hline
		\textbf{Color histograms} & \cite{bedagkar2011multiple} & 2014 & Color histograms could quantify the distribution of colors within images. They prove invaluable in distinguishing various objects and types of land cover present within these images. \\
		\hline
		\textbf{CNN} & \cite{jiang2018geospatial} & 2018 & CNNs are deep learning models particularly effective for image classification tasks in geospatial data analysis. \\
		\hline
		\textbf{IDW} & \cite{varatharajan2018visual} & 2018 & IDW estimates values by giving more weight to nearby observations, assuming spatial similarity. \\
		\hline
		\textbf{Kriging interpolation} & \cite{khan2019geo} & 2019 & Kriging interpolation is an interpolation method that estimates values at unobserved locations by considering spatial autocorrelation in geospatial data. \\
		\hline
		\textbf{RNN} & \cite{cao2019short} & 2019 & RNNs are suitable for time-series data classification in geospatial applications, especially in analyzing temporal changes. \\
		\hline
	\end{tabularx}
\end{table*}

\textbf{Geospatial feature extraction algorithms.}  These algorithms are used to extract feature information from raw geospatial data for subsequent classification and analysis. Common geospatial feature extraction algorithms include edge detection \cite{li2008gis}, texture feature extraction \cite{lucieer2005multivariate}, and color histograms \cite{bedagkar2011multiple}.

\textbf{Deep learning algorithms.} Deep learning algorithms have gained widespread use in recent years for geospatial data classification. Convolutional neural networks (CNN) excel in image data classification \cite{jiang2018geospatial}, while recurrent neural networks (RNN) are suitable for time-series data classification \cite{cao2019short}. Deep learning algorithms can be applied to the feature extraction and classification of geospatial data.

\textbf{Geographic data mining algorithms.} These algorithms excel in the discovery of association rules, identification of spatio-temporal patterns, and more within geographical datasets \cite{shyu2006knowledge}. One notable advantage of these algorithms lies in their capacity for explanation, underlying relationships, and patterns inherent in geographic data.

\textbf{Geospatial interpolation algorithms.} Geospatial interpolation algorithms are used to estimate data values at unknown geographic locations based on existing geospatial data. These algorithms infer missing or unobserved data using available geospatial information. Common geospatial interpolation algorithms include kriging interpolation \cite{khan2019geo} and inverse distance weighting (IDW) \cite{varatharajan2018visual}.

\subsection{Prediction.}

Predictive modeling in geospatial data analysis has a wide range of applications, from forecasting weather patterns and urban growth to predicting disease outbreaks and environmental changes. These algorithms allow researchers and analysts to make decisions and anticipate future developments within a geographic context. By leveraging the power of predictive modeling, valuable insights can be gained to guide planning, resource allocation, and decision-making processes. Here are some common predictive algorithms for geospatial data, as shown in Table \ref{table:2}:

\begin{table*}[ht]
	\caption{The prediction for geospatial data}
	\label{table:2}
	\begin{tabularx}{\textwidth}{|m{3cm}<{\centering}|m{1cm}<{\centering}|m{1cm}<{\centering}|X|}
		\hline
		\textbf{Algorithm} & \textbf{Paper} & \textbf{Year} & \multicolumn{1}{c|}{\textbf{Description}} \\
		\hline
		\textbf{Agent-based modeling} & \cite{macal2005tutorial} & 2005 & Agent-based modeling uses a simulation technique that models individual agents' behavior and interactions within a geographic context. \\
		\hline
		\textbf{Seasonal ARIMA} & \cite{chen2009seasonal} & 2009 & Seasonal ARIMA could extend ARIMA to account for seasonality in time series data, making it suitable for forecasting periodic geospatial phenomena. \\
		\hline
		\textbf{Linear regression} & \cite{thapa2012geographically} & 2012 & Linear regression can predict attribute values at specific geographic locations based on linear relationships between variables. \\
		\hline
		\textbf{SVM} & \cite{pradhan2013comparative} & 2013 & SVM is a machine learning algorithm that can be used for regression tasks to predict attribute values. \\
		\hline
		\textbf{SLM} & \cite{bidanset2014evaluating} & 2014 & SLM can examine how the values of a given variable in one location are influenced by the values of the same variable in neighboring locations. \\
		\hline
		\textbf{ARIMA} & \cite{mehrmolaei2016time} & 2016 & ARIMA is a time series forecasting method that models temporal data patterns and can predict future values based on past observations. \\
		\hline
		\textbf{SAR} &\cite{ver2018spatial} & 2018 & SAR can incorporate spatial autocorrelation and consider interactions among geographical data points to predict attribute values. \\
		\hline
	\end{tabularx}
\end{table*}

\textbf{Geospatial regression prediction.}  Geospatial regression analysis integrates geospatial autocorrelation and attribute relationships to predict attribute values within geospatial data. It considers interactions and correlations among geographical data points. Common approaches are spatial autoregressive models (SAR) \cite{ver2018spatial} and geospatial lag models (SLM) \cite{bidanset2014evaluating}.

\textbf{Time series prediction.} Time series analysis could be used to predict temporal data patterns. In geospatial data, time series analysis can forecast geographic phenomena over time, such as temperature fluctuations or changes in traffic flow over the next few days. Common time series analysis methods include autoregressive integrated moving average (ARIMA) \cite{mehrmolaei2016time} and seasonal ARIMA (SARIMA).

\textbf{Machine learning algorithms.} These algorithms can predict complex relationships and trends within geospatial data. Regression algorithms like linear regression \cite{thapa2012geographically} and support vector machine (SVM) can forecast attribute values at specific geographic locations \cite{pradhan2013comparative}. Deep learning algorithms, such as neural networks, enable more sophisticated modeling and prediction of geospatial data, including object recognition in image data or trend forecasting in time series data.

\textbf{Predictive geographic models.} Predictive geographic models harness geospatial data and geographic features to anticipate future geographic phenomena. These models leverage a variety of techniques that simulate individual agents' behavior and interactions within a geographic context. Agent-based modeling is used for predicting complex geospatial phenomena like urban growth or disease spread \cite{macal2005tutorial}. Moreover, predictive geographic models analyze historical data and the relationships between geospatial elements to offer valuable insights into prospective geographic developments, such as population distribution trends.

\subsection{Visualization.}

Visualizing geospatial data involves a variety of techniques and methods that transform complex geographic information into intuitive graphical representations \cite{resch2014web}. These visualizations are important in facilitating a deeper understanding of geospatial data. These visualization techniques serve as powerful tools for conveying geospatial information, and uncovering hidden patterns within complex geographic data. In this section, we give detailed explanations of several common geospatial visualization techniques.

\textbf{Geospatial scatter plots.} Geospatial scatter plots adapt traditional scatter plots for visualizing spatial distributions and relationships by plotting data points on geographical maps, using latitude and longitude as axes or overlaying points onto a map base. This adaptation enhances scatter plots by incorporating geographical context, allowing visualization of how variables distribute across geographic space by highlighting patterns, clusters, or outliers \cite{anselin2009opengeoda}. 

\textbf{3D visualization techniques.} They are instrumental in creating immersive, three-dimensional models and scenes of geographic spaces \cite{li2022three}. These techniques add depth and realism to geospatial visualizations, providing a more accurate representation of urban structures. By leveraging 3D visualization, users can gain a better understanding of complex geospatial data, enabling them to make decisions and analyze data more intuitively.

\textbf{Virtual reality (VR) and augmented reality (AR).} By integrating geospatial data with virtual or augmented environments, these technologies offer immersive experiences that enable users to interact with geographic information in innovative ways \cite{chen2023open}. VR can provide a fully digital environment that simulates real-world environments, while AR overlays digital information onto the real world, creating a hybrid experience that enhances the perception of reality.

\subsection{Integrated framework.}

\begin{figure}[htbp]
	\centering
	\includegraphics[width=0.48\textwidth]{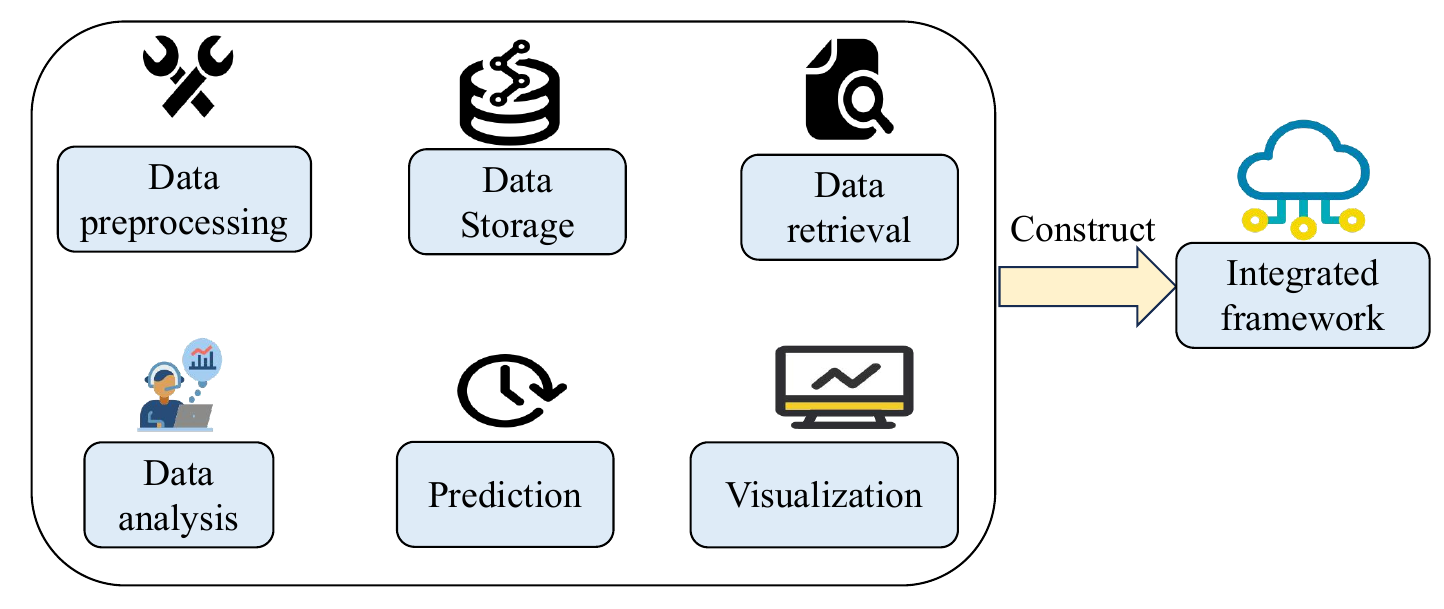} 
	\caption{The combinations with cross technologies.}
	\label{cross} 
\end{figure}

Integrate the techniques introduced above, including data collection, preprocessing, storage, retrieval, analysis, prediction, and visualization, into a unified application scenario \cite{pan20203d}, as shown in Fig. \ref{cross}.

\textbf{Illustrative case.} To demonstrate the practical application of the geospatial data framework, let's consider the task of urban planning for sustainable development. The process begins by gathering geospatial data from various sources, such as satellite imagery, GPS data, and existing maps. These data are then cleaned through preprocessing techniques. Spatial analysis methods are employed to identify the areas for development, taking into account factors such as land use, zoning regulations, and environmental conditions. The predictive models are then used to forecast patterns of population growth and identify areas of high demand for housing and infrastructure. Finally, visualization methods are utilized to effectively present the results to leaders, allowing them to make decisions about sustainable urban development. By leveraging this geospatial framework, decision-makers can create a more livable urban environment.

\textbf{Scalability exploration.}  An essential aspect of the geospatial data framework is its ability to scale, enabling it to handle large-scale data processing and distributed computing. This scalability is particularly crucial in scenarios involving geographic complex data sets, where traditional processing methods may prove inadequate. To address this challenge, techniques such as parallel computing and distributed processing can be leveraged to ensure efficient data analysis and management \cite{rashid2018distributed}. By harnessing these techniques, the geospatial data framework can effectively handle large volumes of data.

\textbf{Incorporating mining techniques.} By incorporating intelligent technologies, such as data processing, data mining, and decision support, the quality of geospatial data processing can be enhanced. For example,  data cleaning can identify errors in the data, minimizing the risk of inaccuracies \cite{chu2015katara}. Data mining techniques can uncover hidden spatial patterns or correlations, extracting the important information from the GBD. Moreover, decision support systems can assist leaders in making well-informed choices based on the analyzed geospatial data.

\section{Advancement on  Processing Technologies: LLM} \label{sec:LLM}

LLM with GBD \cite{zhou2023llm,saeed2023querying} is a potential way for GBD mining in the future. At present, many large models will be used in knowledge-based questions and answers. At the same time, with the help of Text-to-SQL, the GBD can also be retrieved for replies. The intersection of LLM and GBD presents numerous opportunities for enhancing geospatial information processing and natural language understanding \cite{wu2023ai}. By integrating these technologies, we can develop more accurate and effective geospatial-related applications. In this section, we'll explore the intersections between LLM and GBD at the algorithmic level, highlighting four key areas of intersection:

\textbf{Geospatial context understanding.} It could be regarded as the aspect of natural language understanding, as it allows models to associate geographical information with text and provide more accurate responses \cite{janowicz2020geoai}. By integrating GBD with LLM, we can enhance their ability to comprehend location-specific references and deliver relevant information. For instance, a chatbot can utilize GBD to understand the context of a user's query related to a specific location, providing the user with useful information about nearby attractions. This integration enables LLM to provide more personalized and context-aware responses, improving the overall user experience.

\textbf{Geographic location inference.} Geographic location inference is the process of deducing geographic locations that are not explicitly mentioned in text \cite{zhou2012geo}. By leveraging GBD, LLM can infer locations based on contextual clues, such as named entities, landmarks, or other location-related information present in the text. This enables LLM to provide accurate responses even when explicit location references are absent, improving the efficiency of location-based services and applications. With this capability, LLMs can better understand the geographic context of natural language inputs, enhancing their ability to provide relevant and personalized responses.

\textbf{Geographic information generation.} Geographic information generation is a process that involves the creation of natural language descriptions for geographic locations \cite{lunga2020apache}. By combining the language generation capabilities of LLM with GBD, it becomes possible to automatically generate detailed and contextually relevant descriptions of locations. This can be particularly useful in applications such as travel planning, tourism, or real estate, where users often require comprehensive information about specific places.

\textbf{Multimodal analysis.} It combines text data with geospatial imagery and map data to solve complex geographical problems \cite{park2021review}. By integrating GBD with LLM, we can analyze and understand both textual information and visual representations of geospatial data. This integration enables models to perform tasks such as geospatial image captioning, geospatial object recognition, or geospatial data fusion. For instance, a self-driving car's AI system can leverage multimodal analysis to recognize objects in its surroundings, understand traffic patterns, and generate concise descriptions of the environment for navigation purposes. This multimodal approach enhances the accuracy and efficiency of geospatial analysis.

\section{GBD Application Scenario I: Urban Management} \label{sec:cities}

\textbf{Transportation planning.} GBD is important for modern transportation systems, providing insights into traffic patterns, congestion, and travel times within cities through real-time traffic monitoring \cite{zhu2018big}. This data is crucial for optimizing transportation systems, identifying bottlenecks, and enhancing overall mobility. By leveraging GBD, urban planners and transportation officials can design optimal routes, determine strategic locations for public transportation stops, and predict demand for various modes of public transport. This enables cities to create more efficient and effective transportation systems, reducing congestion and improving air quality for their citizens.

\textbf{Infrastructure management.} GBD mining is also important for infrastructure planning, development, and maintenance \cite{arfat2020big}. By analyzing geospatial data related to utilities, buildings, and land use, planners can identify areas needing infrastructure upgrades and allocate resources. This enables cities to prioritize their infrastructure investments and make data-driven decisions about where to focus their efforts. Moreover, geospatial analysis aids in assessing the condition of existing infrastructure, such as roads, bridges, and utilities, enabling proactive maintenance and reducing the risk of failures. This not only improves the efficiency of infrastructure maintenance but also helps to ensure public safety and minimize the impact of potential disruptions.

\textbf{Public services allocation.} GBD mining is a strong tool for making sure public services are distributed well in cities \cite{joseph2013big}. By looking at population data, how land is used, and how easy it is to get to services, leaders can see where more services like healthcare, schools, and public spaces are needed. This makes sure resources go to where they're most needed, ensuring everyone has access to important services. Also, using geospatial analysis helps make smart, evidence-based choices in delivering city services. This leads to services that meet people's needs better and a better life for people living in cities.

\textbf{Emergency response.} GBD mining is useful in emergency response and disaster management within cities \cite{akter2019big}. By integrating data from various sources, such as emergency calls, weather data, and infrastructure maps, emergency services can leverage GBD to assess risks, plan evacuation routes, and allocate resources. This enables emergency responders to react quickly, minimizing the impact of disasters and saving lives. Moreover, geospatial analysis helps identify vulnerable areas, improve disaster preparedness, and coordinate response efforts during emergencies. By using GBD, cities can become stronger against natural and man-made disasters, making a safer and more secure place for people to live.

\section{GBD Application Scenario II: Environmental  Sustainability} \label{sec:enviromental}

\textbf{Urban planning.} GBD mining can be used to offer enhanced quality of life for residents \cite{rathore2016urban}. By optimizing infrastructure and fostering sustainable growth, GBD helps identify suitable construction sites and supports urban planning decisions. This technology also helps check how city building affects the environment, encouraging green practices to make sure cities develop in a way that's good for the planet. In addition, GBD improves the efficient management of essential city systems such as water supply networks, waste management, and energy distribution grids. Geospatial analysis helps find areas that don't have enough services, making cities more welcoming and easy to live in for everyone. By harnessing the power of GBD, cities can become smarter, more sustainable, and better equipped to meet the needs of their residents.

\textbf{Environmental conservation.} Satellite figures and remote sensing data give important information about changes in land, cutting down forests, and the health of ecosystems. This lets scientists and conservationists find areas that are in danger and take steps to protect places with lots of different plants and animals and delicate ecosystems \cite{runting2020opportunities}. By utilizing geospatial data, experts can keep an eye on how human actions affect natural resources. This includes tracking illegal tree cutting, checking the health of coral reefs, and watching how invasive species spread. This information is crucial for designing effective conservation strategies and promoting sustainable resource management. With the help of GBD, it can ensure a healthier, more sustainable future for our planet.

\textbf{Agriculture and food security.}  By leveraging remote sensing technologies and geospatial data, farmers can precisely monitor agricultural land, crop health, and soil conditions, enabling them to optimize irrigation, fertilizer application, and pest control \cite{jin2020big}. This makes farming more productive and resource-efficient, reducing waste and environmental damage. It uses smart farming methods, such as adjusting planting and harvesting amounts based on need. Moreover, GBD enables the comprehensive assessment of food security by monitoring crop production, identifying areas susceptible to food shortages, and facilitating timely interventions to mitigate potential risks. With GBD, it can ensure a more sustainable and secure food supply.

\textbf{Climate change mitigation.} By analyzing geospatial data over a long time, scientists can notice weather changes, like temperature, rain patterns, rising sea levels, and melting glaciers. This information helps in creating climate models and planning how to deal with and reduce the effects of climate change \cite{garrett2022climate}.
GBD also supports the monitoring and verification of greenhouse gas emissions. Moreover, GBD  helps find where we can get energy from renewable sources, choose the best places for wind and solar farms, and plan strong structures to mitigate the impacts of climate change. With GBD, we can better understand and address the issue of climate change.

\section{Open challenges} \label{sec:challenges}

GBD faces several pressing challenges that must be addressed. Firstly, to better utilize the vast amounts of GBD, it's necessary to improve the performance of data retrieval methods. Moreover, privacy concerns are a significant issue, as the use of geospatial data raises sensitive problems about data protection. 

\subsection{Data retrieval issues}

There is a growing demand for retrieval systems that can adapt to user queries and preferences, providing personalized results that match their expectations. This requires efficient data retrieval from GBD, which poses a central challenge for GIS \cite{wiegand2007task}. The primary objective of this optimization is to quickly and accurately locate specific data within the vast amount of geospatial information available. However, this task is challenging due to the vast scale and diverse GBD. A key challenge is creating robust spatial indexes that can handle various types of geospatial data, ranging from simple points to complex polygons, and consider multi-scale data for efficient retrieval. Moreover, ensuring fast query performance with large datasets, especially when dealing with complex spatial queries.

To address these challenges, optimizing indexing structures, query processing algorithms, and distributed computing techniques can help achieve fast and scalable data retrieval. By leveraging advanced indexing techniques, such as spatial indexes, and optimizing query processing algorithms, such as spatial join algorithms, it can significantly reduce the time it takes to retrieve and analyze large volumes of geospatial data. Moreover, distributing data and computing tasks across multiple machines using distributed computing techniques, such as Hadoop or Spark \cite{huang2016memory}, can further enhance performance and scalability. These optimizations enable real-time access to geospatial information, supporting applications that require timely insights and decision-making, such as emergency response, traffic management, or location-based services.

\subsection{Privacy issues}

Geospatial data plays a vital role in various applications. However, it also raises significant security and privacy concerns, particularly when handling sensitive location information \cite{hoh2007preserving}. Safeguarding such data requires careful consideration of handling, storage, and processing practices. To effectively mitigate these risks, it is crucial to implement robust data management policies with a range of protective measures. 

To address the significant security and privacy concerns raised by geospatial data, especially when dealing with sensitive location information, two primary strategies are employed: data encryption \cite{hiemenz2019dynamic} and anomaly detection \cite{budgaga2017framework}. Data encryption is essential for protecting the confidentiality and integrity of geospatial data. By encrypting sensitive location information, data can be transformed into a secure format that is unreadable without the appropriate decryption key. This ensures that even if the data is intercepted or accessed without authorization, the information remains protected and unusable to unauthorized parties. Encryption not only secures data in transit and at rest but also provides a foundational layer of privacy by ensuring that sensitive personal details are not directly exposed. Employing encryption techniques minimizes the risks associated with handling geospatial data, allowing valuable insights to be derived from the data while ensuring the protection of individuals' privacy. Additionally, implementing anomaly detection is useful for filtering out malicious data inputs. Anomaly detection involves analyzing data patterns to identify any deviations from the norm that could indicate fraudulent or harmful activity. By identifying and isolating these anomalies, it is possible to prevent the misuse of geospatial data, further enhancing privacy and security measures.

\section{Future directions} \label{sec:future}

\begin{figure}[htbp]
	\centering
	\includegraphics[width=0.48\textwidth]{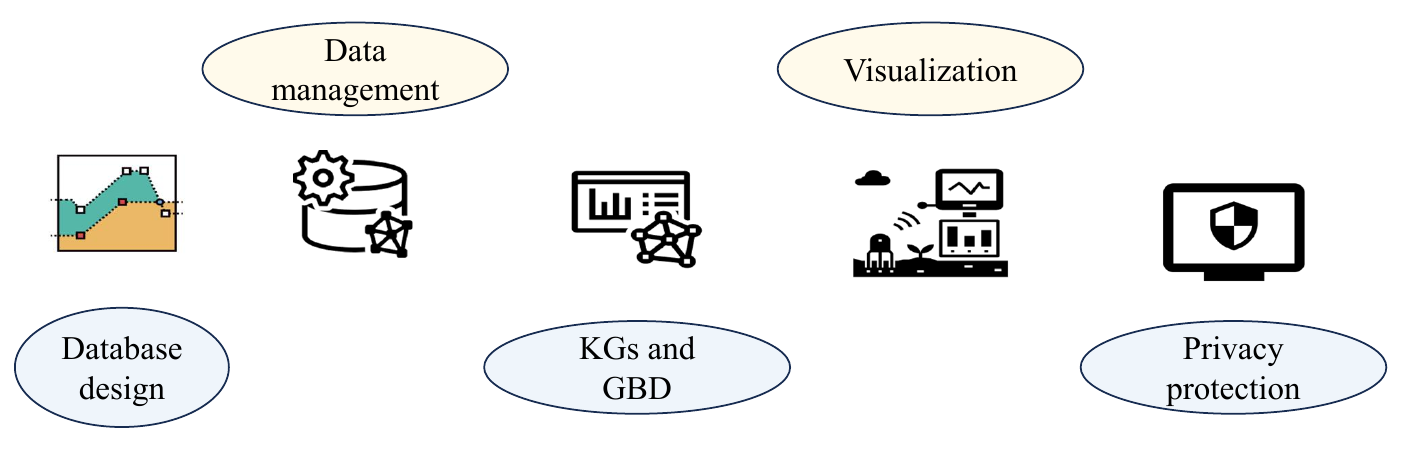} 
	\caption{The future directions.}
	\label{future} 
\end{figure}

The distinctive characteristics of huge data have introduced significant challenges to GBD. GBD must effectively address several critical concerns, as shown in Fig. \ref{future}.

\subsection{Design of GBD-based database}

The design of a GBD-based database involves the vectorized database structure that caters to the unique demands of handling vast and complex geospatial data \cite{bereta2016ontop}. This structure should facilitate efficient storage, retrieval, and manipulation of spatial information, enabling streamlined operations in a big data context.

\textbf{Fault tolerance.} To ensure data integrity and reliability in a GBD-based database, mechanisms for data replication, distribution, and fault tolerance should be incorporated. Replication strategies \cite{agarwal2016performance}, such as master-slave or multi-master replication, can provide redundancy and availability. Data distribution techniques \cite{amiri2021sharper}, like sharding or data partitioning, can enable parallel processing and scalability across multiple nodes or clusters. Fault tolerance mechanisms \cite{huang2016memory}, such as data backups, automated recovery, or distributed consensus algorithms, safeguard the database against failures and ensure continuous availability and disaster recovery.

\textbf{Query optimization.} GBD-based database design involves optimizing query performance to enable fast and efficient retrieval of geospatial information. This includes techniques like query optimization, spatial indexing, caching, and adaptive query processing \cite{christoforaki2011text}. By analyzing query patterns, optimizing data placement, and employing caching strategies, the database can deliver improved query response times and overall performance.

\textbf{Parallel processing.} GBD-based databases should be designed to leverage parallel processing capabilities and scale horizontally to handle GBD 
\cite{dayarathna2018recent}. This involves employing distributed computing techniques, such as parallel query processing, data parallelism, or distributed file systems, to achieve efficient data processing across multiple computing nodes. Scalability can be achieved by adding or removing nodes dynamically to accommodate changing data volumes and processing requirements.

\subsection{Heterogeneous data management}

To enhance reliability, the real-time data management system should possess fault tolerance, ensuring uninterrupted data processing despite node failures or network issues \cite{breunig2020geospatial}. Moreover, it should be designed with scalability in mind, enabling dynamic resource allocation based on data volume fluctuations. Finally, incorporating real-time monitoring capabilities facilitates detection and resolution, ensuring a smooth and responsive data processing workflow. The rapid growth of GBD requires an efficient real-time data management system to handle continuous updates and processing. To achieve this, some robust measures are required:

\textbf{Multimodal data.} The diversity of data presents significant challenges in terms of data harmonization, as each data type may follow different standards, scales, and formats. An effective real-time data management system must be capable of normalizing these data formats to allow for integration and analysis. Semantic alignment helps derive meaningful insights from multimodal data, by ensuring consistency \cite{zhang2013semantic}. Semantic alignment involves matching related information across different datasets. This process is crucial for maintaining the integrity and usability of data.

\textbf{Vector representation.} Vectorizing geospatial data, which involves converting geographic information such as maps, remote sensing images, and geographical features into vector data  \cite{xavier2016survey}, enables the facilitation of GBD mining and learning tasks by representing and processing geographic data in the form of geometric shapes and attribute tables.

\textbf{Stream data.} To accommodate the continuous flow of geospatial data, the system needs to have stream data processing capabilities \cite{wang2019integrated}. This involves the ability to receive, parse, and process data streams in real time. Real-time data management allows for timely updates and analysis, enabling users to make decisions based on the most up-to-date information available. Stream processing frameworks, such as Apache Kafka\footnote{https://kafka.apache.org/} or Apache Flink\footnote{https://flink.apache.org/}, can be utilized to handle GBD streams effectively.

\subsection{Integration with knowledge graphs} Integrating GBD with knowledge graphs provides significant semantic advantages for understanding, querying, and analyzing geospatial information \cite{karalis2019extending}. By integrating geospatial data with knowledge graphs, users can leverage the rich semantic relationships and ontologies present in knowledge graphs to better understand the context of geospatial data. This enables more helpful querying and analysis capabilities, such as spatial reasoning, entity disambiguation, and fusion of multi-source data. Moreover, knowledge graphs can provide a unified representation of geospatial data, facilitating interoperability across different data sources and systems.

\textbf{Semantic modeling.} Semantic modeling involves extracting entities, attributes, and relationships from GBD and mapping them to concepts and relationships in a knowledge graph \cite{liu2021heterogeneous}. This process establishes a richer semantic model for geospatial information, enabling more effective querying and analysis. By capturing the semantics of geospatial data, the system gains a deeper understanding of the underlying information and can provide more accurate and meaningful information.

\textbf{Intelligent querying.} The semantic information stored in the knowledge graph enables intelligent querying \cite{dsouza2021worldkg}. Users can now interact with the system using natural language or complex query statements. The system comprehends the query intents, understands the context, and can provide more accurate and comprehensive results. This capability enhances the user experience and enables more intuitive exploration and analysis of geospatial data.

\textbf{Geospatial reasoning.} Leveraging the semantic information from the knowledge graph allows for association mining on GBD. This process helps identify implicit relationships and patterns that may not be immediately apparent. By discovering hidden connections, the system can infer meaningful associations. The reasoning capabilities of the knowledge graph improve geospatial reasoning. By analyzing GBD in the context of the knowledge graph, the system can uncover latent connections and influencing factors among the data. 

\subsection{Visualization}

The evolution of GIS is steering towards an enhanced visualization paradigm by leveraging advanced 3D technologies and the concept of the Metaverse. 

\textbf{Visualization enhancement.} Modern techniques in 3D modeling, rendering, and animation help create interactive and lifelike visuals. This allows us to see patterns that are not immediately obvious and explore different places and environments in a detailed and straightforward way. It allows the creation of immersive and lifelike representations of spatial data, enabling a deeper understanding of complex geographical contexts \cite{moser2010beyond}. 

\textbf{Metaverse integration.} The emerging concept of the Metaverse \cite{sun2022metaverse} further amplifies GIS capabilities with the combination of real-world geographic data and virtual environments \cite{chen2023open}. This shared digital application enables tangible interactions with geospatial information, fostering collaboration and decision-making. By using the Metaverse, GIS visualization gains an immersive and interactive dimension, offering novel perspectives for urban planning, disaster management, environmental monitoring, and other domains. The Metaverse provides a platform where leaders can engage with spatial data in dynamic and transformative ways.

\subsection{Privacy protection}

Privacy protection is a primary consideration when processing and analyzing geospatial data. In today's digital society, the security and privacy of personal data have become a central point of concern for the public and professional institutions. The development of technological methods, especially anomaly detection, federated learning, and differential privacy, offers new pathways and methods to address these issues. These technologies can effectively identify and guard against potential data leaks and misuse while allowing for the collection and analysis of geospatial data without compromising individual privacy. This is crucial for urban planning, environmental monitoring, public health, and many other fields. By enabling intelligent processing and analysis of data, we can extract valuable insights and information without sacrificing personal privacy.

\textbf{Anomaly detection.} It aims to identify data points, events, or observations that significantly deviate from the dataset's normal behavior, indicating potential issues such as data corruption, fraudulent activities, or environmental anomalies. For this purpose, methods such as normal distribution analysis and contrastive learning can be utilized. Normal distribution analysis assumes that the majority of the data follows a Gaussian distribution \cite{jiang2015geospatial}, considering data points that lie several standard deviations away from the mean as anomalies. This method is suitable for datasets with a well-defined normal state but requires precise adjustment to distinguish between true anomalies and natural variability. Contrastive learning, as a form of self-supervised learning, identifies instances that significantly differ from the normal data representation by training models to recognize pairs of data that are normal and anomalous \cite{mai2023csp,yang2022sparse}. This approach is particularly effective in handling complex geospatial data, allowing models to learn to identify anomalous patterns without explicit anomaly labels. This is especially valuable in scenarios where anomalies cannot be directly defined through simple statistical metrics. Through this self-supervised learning process, models can leverage the structure of the data, enhancing their ability to recognize anomalies and playing a crucial role in protecting data integrity and user privacy.

\textbf{Federated learning.} Federated learning represents a paradigm shift in how geospatial data is processed and analyzed, particularly in terms of privacy preservation \cite{tam2021adaptive, sprague2018asynchronous}. By allowing data analysis models to be trained locally on devices, and then aggregating the model updates rather than the raw data, federated learning minimizes the exposure of sensitive information. This decentralized approach ensures that personal and geographically identifiable data does not leave its original context, significantly reducing the risk of privacy leaks. Furthermore, federated learning can be optimized for geospatial data by incorporating spatial hierarchies and dependencies into the learning process, enhancing model accuracy and utility without compromising privacy. 

\textbf{Differential privacy.} It is a vital component of protecting individual privacy when working with geospatial data \cite{algarni2023p3s}. By introducing controlled noise into the data, differential privacy ensures that no single data point can significantly impact analysis outcomes, preserving individual identities. This innovative approach enables researchers and analysts to extract valuable insights from geospatial data while maintaining the privacy of individuals. By safeguarding sensitive information, differential privacy fosters trust in data-driven decision-making and allows for the responsible use of geospatial data in various applications.

\section{Conclusions} \label{sec:conclusion}

In summary, this paper surveys Geospatial Big Data (GBD) and discusses the latest developments, how it's used, the challenges faced, and what the future might hold. It demonstrates how GBD mining works in a unified framework and can be even more powerful for applications like urban management and environmental sustainability. Although GBD has a lot of potential, there are also big challenges like how to improve data retrieval and protect data privacy. This paper suggests several ways to move forward, including optimizing the framework to manage the data, using knowledge graphs to connect different pieces of information more effectively, making data visualization more immersive, and using effective ways to keep data secure. Overall, this research offers a clear picture of where GBD stands today and gives directions on how it could evolve to better serve our communities in the future.

\bibliographystyle{IEEEtran}
\bibliography{geospatialdata}

\begin{thebibliography}{100}
\providecommand{\url}[1]{#1}
\csname url@samestyle\endcsname
\providecommand{\newblock}{\relax}
\providecommand{\bibinfo}[2]{#2}
\providecommand{\BIBentrySTDinterwordspacing}{\spaceskip=0pt\relax}
\providecommand{\BIBentryALTinterwordstretchfactor}{4}
\providecommand{\BIBentryALTinterwordspacing}{\spaceskip=\fontdimen2\font plus
\BIBentryALTinterwordstretchfactor\fontdimen3\font minus
  \fontdimen4\font\relax}
\providecommand{\BIBforeignlanguage}[2]{{%
\expandafter\ifx\csname l@#1\endcsname\relax
\typeout{** WARNING: IEEEtran.bst: No hyphenation pattern has been}%
\typeout{** loaded for the language `#1'. Using the pattern for}%
\typeout{** the default language instead.}%
\else
\language=\csname l@#1\endcsname
\fi
#2}}
\providecommand{\BIBdecl}{\relax}
\BIBdecl

\bibitem{gan2023web}
W.~Gan, Z.~Ye, S.~Wan, and P.~S. Yu, ``Web 3.0: The future of internet,'' in
  \emph{Companion Proceedings of the ACM Web Conference}, 2023, pp. 1266--1275.

\bibitem{wan2023web3}
S.~Wan, H.~Lin, W.~Gan, J.~Chen, and P.~S. Yu, ``Web3: The next internet
  revolution,'' \emph{arXiv preprint arXiv:2304.06111}, 2023.

\bibitem{lee2015geospatial}
J.-G. Lee and M.~Kang, ``Geospatial big data: challenges and opportunities,''
  \emph{Big Data Research}, vol.~2, no.~2, pp. 74--81, 2015.

\bibitem{lee2011recent}
C.~A. Lee, S.~D. Gasster, A.~Plaza, C.-I. Chang, and B.~Huang, ``Recent
  developments in high performance computing for remote sensing: A review,''
  \emph{IEEE Journal of Selected Topics in Applied Earth Observations and
  Remote Sensing}, vol.~4, no.~3, pp. 508--527, 2011.

\bibitem{brynjolfsson2017artificial}
E.~Brynjolfsson and A.~Mcafee, ``Artificial intelligence, for real,''
  \emph{Harvard business review}, vol.~1, pp. 1--31, 2017.

\bibitem{shekhar2010spatial}
S.~Shekhar, P.~Zhang, and Y.~Huang, ``Spatial data mining,'' \emph{Data Mining
  and Knowledge Discovery Handbook}, pp. 837--854, 2010.

\bibitem{wang2020deep}
S.~Wang, J.~Cao, and P.~S. Yu, ``Deep learning for spatio-temporal data mining:
  A survey,'' \emph{IEEE Transactions on Knowledge and Data Engineering},
  vol.~34, no.~8, pp. 3681--3700, 2020.

\bibitem{mennis2009spatial}
J.~Mennis and D.~Guo, ``Spatial data mining and geographic knowledge
  discovery—an introduction,'' \emph{Computers, Environment and Urban
  Systems}, vol.~33, no.~6, pp. 403--408, 2009.

\bibitem{lu2003algorithms}
C.-T. Lu, D.~Chen, and Y.~Kou, ``Algorithms for spatial outlier detection,'' in
  \emph{the third International Conference on Data Mining}.\hskip 1em plus
  0.5em minus 0.4em\relax IEEE, 2003, pp. 597--600.

\bibitem{keim2004visual}
D.~A. Keim, C.~Panse, M.~Sips, and S.~C. North, ``Visual data mining in large
  geospatial point sets,'' \emph{IEEE Computer Graphics and Applications},
  vol.~24, no.~5, pp. 36--44, 2004.

\bibitem{zhao2021deep}
B.~Zhao, S.~Zhang, C.~Xu, Y.~Sun, and C.~Deng, ``Deep fake geography? when
  geospatial data encounter artificial intelligence,'' \emph{Cartography and
  Geographic Information Science}, vol.~48, no.~4, pp. 338--352, 2021.

\bibitem{wu2023multimodal}
J.~Wu, W.~Gan, Z.~Chen, S.~Wan, and P.~S. Yu, ``Multimodal large language
  models: A survey,'' in \emph{IEEE International Conference on Big
  Data}.\hskip 1em plus 0.5em minus 0.4em\relax IEEE, 2023, pp. 2247--2256.

\bibitem{han2023chartllama}
Y.~Han, C.~Zhang, X.~Chen, X.~Yang, Z.~Wang, G.~Yu, B.~Fu, and H.~Zhang,
  ``{ChartLlama}: A multimodal {LLM} for chart understanding and generation,''
  \emph{arXiv preprint arXiv:2311.16483}, 2023.

\bibitem{mai2022symbolic}
G.~Mai, Y.~Hu, S.~Gao, L.~Cai, B.~Martins, J.~Scholz, J.~Gao, and K.~Janowicz,
  ``Symbolic and subsymbolic geoai: Geospatial knowledge graphs and spatially
  explicit machine learning.'' \emph{Trans. GIS}, vol.~26, no.~8, pp.
  3118--3124, 2022.

\bibitem{zauskova2022visual}
A.~Zauskova, R.~Miklencicova, and G.~H. Popescu, ``Visual imagery and
  geospatial mapping tools, virtual simulation algorithms, and deep
  learning-based sensing technologies in the metaverse interactive
  environment,'' \emph{Review of Contemporary Philosophy}, vol.~21, pp.
  122--137, 2022.

\bibitem{jiang2010geospatial}
B.~Jiang and X.~Yao, ``Geospatial analysis and modeling of urban structure and
  dynamics: An overview,'' \emph{Geospatial Analysis and Modelling of Urban
  Structure and Dynamics}, pp. 3--11, 2010.

\bibitem{rai2022geospatial}
P.~K. Rai, V.~N. Mishra, and P.~Singh, \emph{Geospatial technology for
  landscape and environmental management: sustainable assessment and
  planning}.\hskip 1em plus 0.5em minus 0.4em\relax Springer, 2022.

\bibitem{shen2012geospatial}
Z.~Shen, \emph{Geospatial techniques in urban planning}.\hskip 1em plus 0.5em
  minus 0.4em\relax Springer Science \& Business Media, 2012.

\bibitem{jiang2018geospatial}
W.~Jiang and L.~Zhang, ``Geospatial data to images: A deep-learning framework
  for traffic forecasting,'' \emph{Tsinghua Science and Technology}, vol.~24,
  no.~1, pp. 52--64, 2018.

\bibitem{brunner2009distributed}
D.~Brunner, G.~Lemoine, F.-X. Thoorens, and L.~Bruzzone, ``Distributed
  geospatial data processing functionality to support collaborative and rapid
  emergency response,'' \emph{IEEE Journal of Selected Topics in Applied Earth
  Observations and Remote Sensing}, vol.~2, no.~1, pp. 33--46, 2009.

\bibitem{avtar2020utilizing}
R.~Avtar, R.~Aggarwal, A.~Kharrazi, P.~Kumar, and T.~A. Kurniawan, ``Utilizing
  geospatial information to implement sdgs and monitor their progress,''
  \emph{Environmental Monitoring and Assessment}, vol. 192, pp. 1--21, 2020.

\bibitem{lacroix2019mapx}
P.~Lacroix, F.~Moser, A.~Benvenuti, T.~Piller, D.~Jensen, I.~Petersen,
  M.~Planque, and N.~Ray, ``{MapX}: An open geospatial platform to manage,
  analyze and visualize data on natural resources and the environment,''
  \emph{SoftwareX}, vol.~9, pp. 77--84, 2019.

\bibitem{hong2007experimental}
Y.~Hong, R.~F. Adler, and G.~Huffman, ``An experimental global prediction
  system for rainfall-triggered landslides using satellite remote sensing and
  geospatial datasets,'' \emph{IEEE Transactions on Geoscience and Remote
  Sensing}, vol.~45, no.~6, pp. 1671--1680, 2007.

\bibitem{hu2016resource}
X.~Hu, X.~Zhang, and J.~Qu, ``Resource storage and management method of massive
  remote sensing data supported by the big data architecture,'' \emph{Journal
  of Geo-information Science}, vol.~18, pp. 681--689, 2016.

\bibitem{li2017geo}
S.~LI, H.~YANG, Y.~HUANG, and Q.~ZHOU, ``Geo-spatial big data storage based on
  nosql database,'' \emph{Geomatics and Information Science of Wuhan
  University}, vol.~42, no.~2, pp. 163--169, 2017.

\bibitem{al2021review}
S.~Al-Yadumi, T.~E. Xion, S.~G.~W. Wei, and P.~Boursier, ``Review on
  integrating geospatial big datasets and open research issues,'' \emph{IEEE
  Access}, vol.~9, pp. 10\,604--10\,620, 2021.

\bibitem{bertino2008security}
E.~Bertino, B.~Thuraisingham, M.~Gertz, and M.~L. Damiani, ``Security and
  privacy for geospatial data: concepts and research directions,'' in \emph{the
  International Workshop on Security and Privacy in GIS and LBS}, 2008, pp.
  6--19.

\bibitem{triglav2011spatio}
J.~Triglav, D.~Petrovi{\v{c}}, and B.~Stopar, ``Spatio-temporal evaluation
  matrices for geospatial data,'' \emph{International Journal of Applied Earth
  Observation and Geoinformation}, vol.~13, no.~1, pp. 100--109, 2011.

\bibitem{anselin2009geoda}
L.~Anselin, I.~Syabri, and Y.~Kho, ``{GeoDa}: an introduction to spatial data
  analysis,'' in \emph{Handbook of Applied Spatial Analysis: Software tools,
  Methods and Applications}.\hskip 1em plus 0.5em minus 0.4em\relax Springer,
  2009, pp. 73--89.

\bibitem{murayama2012progress}
Y.~Murayama, \emph{Progress in geospatial analysis}.\hskip 1em plus 0.5em minus
  0.4em\relax Springer Science \& Business Media, 2012.

\bibitem{de2007geospatial}
M.~J. De~Smith, M.~F. Goodchild, and P.~Longley, \emph{Geospatial analysis: a
  comprehensive guide to principles, techniques and software tools}.\hskip 1em
  plus 0.5em minus 0.4em\relax Troubador publishing ltd, 2007.

\bibitem{goodchild2009geographic}
M.~F. Goodchild, ``Geographic information systems and science: today and
  tomorrow,'' \emph{Annals of GIS}, vol.~15, no.~1, pp. 3--9, 2009.

\bibitem{dold2017future}
J.~Dold and J.~Groopman, ``The future of geospatial intelligence,''
  \emph{Geo-spatial Information Science}, vol.~20, no.~2, pp. 151--162, 2017.

\bibitem{aggarwal2004principles}
S.~Aggarwal, ``Principles of remote sensing,'' \emph{Satellite Remote Sensing
  and GIS Applications in Agricultural Meteorology}, vol.~23, no.~2, pp.
  23--28, 2004.

\bibitem{kursinski1997observing}
E.~Kursinski, G.~Hajj, J.~Schofield, R.~Linfield, and K.~R. Hardy, ``Observing
  earth's atmosphere with radio occultation measurements using the global
  positioning system,'' \emph{Journal of Geophysical Research: Atmospheres},
  vol. 102, no. D19, pp. 23\,429--23\,465, 1997.

\bibitem{monteiro2018urban}
C.~S. Monteiro, C.~Costa, A.~Pina, M.~Y. Santos, and P.~Ferr{\~a}o, ``An urban
  building database ({UBD}) supporting a smart city information system,''
  \emph{Energy and Buildings}, vol. 158, pp. 244--260, 2018.

\bibitem{dodge2008power}
M.~Dodge, M.~McDerby, and M.~Turner, ``The power of geographical
  visualizations,'' \emph{Geographic Visualization}, pp. 1--9, 2008.

\bibitem{ziliaskopoulos2000internet}
A.~K. Ziliaskopoulos and S.~T. Waller, ``An internet-based geographic
  information system that integrates data, models and users for transportation
  applications,'' \emph{Emerging Technologies}, vol.~8, no. 1-6, pp. 427--444,
  2000.

\bibitem{tsou2004integrated}
M.-H. Tsou, ``Integrated mobile gis and wireless internet map servers for
  environmental monitoring and management,'' \emph{Cartography and Geographic
  Information Science}, vol.~31, no.~3, pp. 153--165, 2004.

\bibitem{burrough2015principles}
P.~A. Burrough, R.~A. McDonnell, and C.~D. Lloyd, \emph{Principles of
  geographical information systems}.\hskip 1em plus 0.5em minus 0.4em\relax
  Oxford University Press, USA, 2015.

\bibitem{cura2017scalable}
R.~Cura, J.~Perret, and N.~Paparoditis, ``A scalable and multi-purpose point
  cloud server ({PCS}) for easier and faster point cloud data management and
  processing,'' \emph{ISPRS Journal of Photogrammetry and Remote Sensing}, vol.
  127, pp. 39--56, 2017.

\bibitem{han2017review}
X.-F. Han, J.~S. Jin, M.-J. Wang, W.~Jiang, L.~Gao, and L.~Xiao, ``A review of
  algorithms for filtering the 3d point cloud,'' \emph{Signal Processing: Image
  Communication}, vol.~57, pp. 103--112, 2017.

\bibitem{brisaboa2017efficiently}
N.~R. Brisaboa, G.~d. Bernardo, G.~Guti{\'e}rrez, M.~R. Luaces, and J.~R.
  Param{\'a}, ``Efficiently querying vector and raster data,'' \emph{The
  Computer Journal}, vol.~60, no.~9, pp. 1395--1413, 2017.

\bibitem{xavier2016survey}
E.~M. Xavier, F.~J. Ariza-L{\'o}pez, and M.~A. Urena-Camara, ``A survey of
  measures and methods for matching geospatial vector datasets,'' \emph{ACM
  Computing Surveys}, vol.~49, no.~2, pp. 1--34, 2016.

\bibitem{geman1996active}
D.~Geman and B.~Jedynak, ``An active testing model for tracking roads in
  satellite images,'' \emph{IEEE Transactions on Pattern Analysis and Machine
  Intelligence}, vol.~18, no.~1, pp. 1--14, 1996.

\bibitem{sarker2021mobile}
I.~H. Sarker, M.~M. Hoque, M.~K. Uddin, and T.~Alsanoosy, ``Mobile data science
  and intelligent {APPs}: concepts, {AI}-based modeling and research
  directions,'' \emph{Mobile Networks and Applications}, vol.~26, no.~1, pp.
  285--303, 2021.

\bibitem{naaman2011geographic}
M.~Naaman, ``Geographic information from georeferenced social media data,''
  \emph{SIGSPATIAL Special}, vol.~3, no.~2, pp. 54--61, 2011.

\bibitem{fonte2017assessing}
C.~C. Fonte, V.~Antoniou, L.~Bastin, J.~Estima, J.~J. Arsanjani, J.-C.~L.
  Bayas, L.~See, and R.~Vatseva, ``Assessing vgi data quality,'' \emph{Mapping
  and the Citizen Sensor}, pp. 137--163, 2017.

\bibitem{raup2007glims}
B.~Raup, A.~Racoviteanu, S.~J.~S. Khalsa, C.~Helm, R.~Armstrong, and Y.~Arnaud,
  ``The {GLIMS} geospatial glacier database: A new tool for studying glacier
  change,'' \emph{Global and Planetary Change}, vol.~56, no. 1-2, pp. 101--110,
  2007.

\bibitem{blazquez2018big}
D.~Blazquez and J.~Domenech, ``Big data sources and methods for social and
  economic analyses,'' \emph{Technological Forecasting and Social Change}, vol.
  130, pp. 99--113, 2018.

\bibitem{lunga2020apache}
D.~Lunga, J.~Gerrand, L.~Yang, C.~Layton, and R.~Stewart, ``Apache spark
  accelerated deep learning inference for large scale satellite image
  analytics,'' \emph{the Journal of Selected Topics in Applied Earth
  Observations and Remote Sensing}, vol.~13, pp. 271--283, 2020.

\bibitem{zhang2016multisource}
Y.~Zhang and S.~Prasad, ``Multisource geospatial data fusion via local joint
  sparse representation,'' \emph{the Transactions on Geoscience and Remote
  Sensing}, vol.~54, no.~6, pp. 3265--3276, 2016.

\bibitem{goyal2020grid}
P.~Goyal, J.~S. Challa, D.~Kumar, A.~Bhat, S.~Balasubramaniam, and N.~Goyal,
  ``Grid-r-tree: a data structure for efficient neighborhood and nearest
  neighbor queries in data mining,'' \emph{International Journal of Data
  Science and Analytics}, vol.~10, no.~1, pp. 25--47, 2020.

\bibitem{mathew2010quad}
R.~Mathew and D.~S. Taubman, ``Quad-tree motion modeling with leaf merging,''
  \emph{IEEE Transactions on Circuits and Systems for Video Technology},
  vol.~20, no.~10, pp. 1331--1345, 2010.

\bibitem{baruffa2019comparison}
G.~Baruffa, M.~Femminella, M.~Pergolesi, and G.~Reali, ``Comparison of mongodb
  and cassandra databases for spectrum monitoring as-a-service,'' \emph{IEEE
  Transactions on Network and Service Management}, vol.~17, no.~1, pp.
  346--360, 2019.

\bibitem{gorelick2017google}
N.~Gorelick, M.~Hancher, M.~Dixon, S.~Ilyushchenko, D.~Thau, and R.~Moore,
  ``Google earth engine: Planetary-scale geospatial analysis for everyone,''
  \emph{Remote Sensing of Environment}, vol. 202, pp. 18--27, 2017.

\bibitem{shankar2020serverless}
V.~Shankar, K.~Krauth, K.~Vodrahalli, Q.~Pu, B.~Recht, I.~Stoica,
  J.~Ragan-Kelley, E.~Jonas, and S.~Venkataraman, ``Serverless linear
  algebra,'' in \emph{The 11th ACM Symposium on Cloud Computing}, 2020, pp.
  281--295.

\bibitem{teng2010medical}
C.-C. Teng, J.~Mitchell, C.~Walker, A.~Swan, C.~Davila, D.~Howard, and
  T.~Needham, ``A medical image archive solution in the cloud,'' in \emph{IEEE
  International Conference on Software Engineering and Service Sciences}.\hskip
  1em plus 0.5em minus 0.4em\relax IEEE, 2010, pp. 431--434.

\bibitem{lv2019bim}
Z.~Lv, X.~Li, H.~Lv, and W.~Xiu, ``Bim big data storage in webvrgis,''
  \emph{IEEE Transactions on Industrial Informatics}, vol.~16, no.~4, pp.
  2566--2573, 2019.

\bibitem{agarwal2016performance}
S.~Agarwal and K.~Rajan, ``Performance analysis of {MongoDB} versus
  {PostGIS}/{PostGreSQL} databases for line intersection and point containment
  spatial queries,'' \emph{Spatial Information Research}, vol.~24, pp.
  671--677, 2016.

\bibitem{george2007spatio}
B.~George, S.~Kim, and S.~Shekhar, ``Spatio-temporal network databases and
  routing algorithms: A summary of results,'' in \emph{the 10th International
  Symposium}.\hskip 1em plus 0.5em minus 0.4em\relax Springer, 2007, pp.
  460--477.

\bibitem{papoutsis2020insar}
I.~Papoutsis, C.~Kontoes, S.~Alatza, A.~Apostolakis, and C.~Loupasakis, ``Insar
  greece with parallelized persistent scatterer interferometry: A national
  ground motion service for big copernicus sentinel-1 data,'' \emph{Remote
  Sensing}, vol.~12, no.~19, p. 3207, 2020.

\bibitem{khelifati2023tsm}
A.~Khelifati, M.~Khayati, A.~Dign{\"o}s, D.~Difallah, and P.~Cudr{\'e}-Mauroux,
  ``Tsm-bench: Benchmarking time series database systems for monitoring
  applications,'' \emph{the VLDB Endowment}, vol.~16, no.~11, pp. 3363--3376,
  2023.

\bibitem{walker2001wavelet}
J.~S. Walker and T.~Q. Nguyen, ``Wavelet-based image compression,''
  \emph{Sub-chapter of CRC Press book: Transforms and Data Compression}, 2001.

\bibitem{siddiquie2011image}
B.~Siddiquie, R.~S. Feris, and L.~S. Davis, ``Image ranking and retrieval based
  on multi-attribute queries,'' in \emph{CVPR}.\hskip 1em plus 0.5em minus
  0.4em\relax IEEE, 2011, pp. 801--808.

\bibitem{punitha2005invariant}
P.~Punitha and D.~Guru, ``An invariant scheme for exact match retrieval of
  symbolic images: Triangular spatial relationship based approach,''
  \emph{Pattern Recognition Letters}, vol.~26, no.~7, pp. 893--907, 2005.

\bibitem{chu2021long}
D.~Chu, H.~Shen, X.~Guan, J.~M. Chen, X.~Li, J.~Li, and L.~Zhang, ``Long
  time-series ndvi reconstruction in cloud-prone regions via spatio-temporal
  tensor completion,'' \emph{Remote Sensing of Environment}, vol. 264, p.
  112632, 2021.

\bibitem{jang2022deep}
Y.~K. Jang, G.~Gu, B.~Ko, I.~Kang, and N.~I. Cho, ``Deep hash distillation for
  image retrieval,'' in \emph{European Conference on Computer Vision}.\hskip
  1em plus 0.5em minus 0.4em\relax Springer, 2022, pp. 354--371.

\bibitem{shahabi2010geodec}
C.~Shahabi, F.~Banaei-Kashani, A.~Khoshgozaran, L.~Nocera, and S.~Xing,
  ``Geodec: A framework to effectively visualize and query geospatial data for
  decision-making,'' \emph{IEEE MultiMedia}, 2010.

\bibitem{li2008gis}
Y.~Li, C.~M. Onasch, and Y.~Guo, ``Gis-based detection of grain boundaries,''
  \emph{Journal of Structural Geology}, vol.~30, no.~4, pp. 431--443, 2008.

\bibitem{lucieer2005multivariate}
A.~Lucieer, A.~Stein, and P.~Fisher, ``Multivariate texture-based segmentation
  of remotely sensed imagery for extraction of objects and their uncertainty,''
  \emph{International Journal of Remote Sensing}, vol.~26, no.~14, pp.
  2917--2936, 2005.

\bibitem{bedagkar2011multiple}
A.~Bedagkar-Gala and S.~K. Shah, ``Multiple person re-identification using part
  based spatio-temporal color appearance model,'' in \emph{the IEEE
  International Conference on Computer Vision Workshops}.\hskip 1em plus 0.5em
  minus 0.4em\relax IEEE, 2011, pp. 1721--1728.

\bibitem{cao2019short}
C.~Cao, S.~Dragi{\'c}evi{\'c}, and S.~Li, ``Short-term forecasting of land use
  change using recurrent neural network models,'' \emph{Sustainability},
  vol.~11, no.~19, p. 5376, 2019.

\bibitem{shyu2006knowledge}
C.-R. Shyu, M.~Klaric, G.~Scott, and W.~Mahamaneerat, ``Knowledge discovery by
  mining association rules and temporal-spatial information from large-scale
  geospatial image databases,'' in \emph{IEEE International Symposium on
  Geoscience and Remote Sensing}.\hskip 1em plus 0.5em minus 0.4em\relax IEEE,
  2006, pp. 17--20.

\bibitem{khan2019geo}
A.~R. Khan, M.~Rafique, S.~U. Rahman, M.~Basharat, C.~Shahzadi, and I.~Ahmed,
  ``Geo-spatial analysis of radon in spring and well water using kriging
  interpolation method,'' \emph{Water Supply}, vol.~19, no.~1, pp. 222--235,
  2019.

\bibitem{varatharajan2018visual}
R.~Varatharajan, G.~Manogaran, M.~Priyan, V.~E. Bala{\c{s}}, and C.~Barna,
  ``Visual analysis of geospatial habitat suitability model based on inverse
  distance weighting with paired comparison analysis,'' \emph{Multimedia Tools
  and Applications}, vol.~77, pp. 17\,573--17\,593, 2018.

\bibitem{ver2018spatial}
J.~M. Ver~Hoef, E.~E. Peterson, M.~B. Hooten, E.~M. Hanks, and M.-J. Fortin,
  ``Spatial autoregressive models for statistical inference from ecological
  data,'' \emph{Ecological Monographs}, vol.~88, no.~1, pp. 36--59, 2018.

\bibitem{bidanset2014evaluating}
P.~E. Bidanset and J.~R. Lombard, ``Evaluating spatial model accuracy in mass
  real estate appraisal: A comparison of geographically weighted regression and
  the spatial lag model,'' \emph{Cityscape}, vol.~16, no.~3, pp. 169--182,
  2014.

\bibitem{mehrmolaei2016time}
S.~Mehrmolaei and M.~R. Keyvanpour, ``Time series forecasting using improved
  arima,'' in \emph{the Conference on Artificial Intelligence and
  Robotics}.\hskip 1em plus 0.5em minus 0.4em\relax IEEE, 2016, pp. 92--97.

\bibitem{chen2009seasonal}
C.-F. Chen, Y.-H. Chang, and Y.-W. Chang, ``Seasonal arima forecasting of
  inbound air travel arrivals to taiwan,'' \emph{Transportmetrica}, vol.~5,
  no.~2, pp. 125--140, 2009.

\bibitem{thapa2012geographically}
R.~B. Thapa and R.~C. Estoque, ``Geographically weighted regression in
  geospatial analysis,'' in \emph{Progress in Geospatial Analysis}.\hskip 1em
  plus 0.5em minus 0.4em\relax Springer, 2012, pp. 85--96.

\bibitem{pradhan2013comparative}
B.~Pradhan, ``A comparative study on the predictive ability of the decision
  tree, support vector machine and neuro-fuzzy models in landslide
  susceptibility mapping using gis,'' \emph{Computers \& Geosciences}, vol.~51,
  pp. 350--365, 2013.

\bibitem{macal2005tutorial}
C.~M. Macal and M.~J. North, ``Tutorial on agent-based modeling and
  simulation,'' in \emph{the Winter Simulation Conference}.\hskip 1em plus
  0.5em minus 0.4em\relax IEEE, 2005, pp. 14--pp.

\bibitem{resch2014web}
B.~Resch, R.~Wohlfahrt, and C.~Wosniok, ``Web-based {4D} visualization of
  marine geo-data using webgl,'' \emph{Cartography and Geographic Information
  Science}, vol.~41, no.~3, pp. 235--247, 2014.

\bibitem{anselin2009opengeoda}
L.~Anselin and M.~McCann, ``Opengeoda, open source software for the exploration
  and visualization of geospatial data,'' in \emph{the 17th ACM International
  Conference on Advances in Geographic Information Systems}, 2009, pp.
  550--551.

\bibitem{li2022three}
W.~Li, J.~Zhu, J.-H. Haunert, L.~Fu, Q.~Zhu, and Y.~Dehbi, ``Three-dimensional
  virtual representation for the whole process of dam-break floods from a
  geospatial storytelling perspective,'' \emph{International Journal of Digital
  Earth}, vol.~15, no.~1, pp. 1637--1656, 2022.

\bibitem{chen2023open}
Z.~Chen, W.~Gan, J.~Sun, J.~Wu, and P.~S. Yu, ``Open metaverse: Issues,
  evolution, and future,'' \emph{arXiv preprint arXiv:2304.13931}, 2023.

\bibitem{pan20203d}
D.~Pan, Z.~Xu, X.~Lu, L.~Zhou, and H.~Li, ``{3D} scene and geological modeling
  using integrated multi-source spatial data: Methodology, challenges, and
  suggestions,'' \emph{Tunnelling and Underground Space Technology}, vol. 100,
  p. 103393, 2020.

\bibitem{rashid2018distributed}
Z.~N. Rashid, S.~R. Zebari, K.~H. Sharif, and K.~Jacksi, ``Distributed cloud
  computing and distributed parallel computing: A review,'' in \emph{the
  International Conference on Advanced Science and Engineering}.\hskip 1em plus
  0.5em minus 0.4em\relax IEEE, 2018, pp. 167--172.

\bibitem{chu2015katara}
X.~Chu, J.~Morcos, I.~F. Ilyas, M.~Ouzzani, P.~Papotti, N.~Tang, and Y.~Ye,
  ``Katara: A data cleaning system powered by knowledge bases and
  crowdsourcing,'' in \emph{the ACM SIGMOD International Conference on
  Management of Data}, 2015, pp. 1247--1261.

\bibitem{zhou2023llm}
X.~Zhou, G.~Li, and Z.~Liu, ``{LLM} as {DBA},'' \emph{arXiv preprint
  arXiv:2308.05481}, 2023.

\bibitem{saeed2023querying}
M.~Saeed, N.~De~Cao, and P.~Papotti, ``Querying large language models with
  {SQL},'' \emph{arXiv preprint arXiv:2304.00472}, 2023.

\bibitem{wu2023ai}
J.~Wu, W.~Gan, Z.~Chen, S.~Wan, and H.~Lin, ``Ai-generated content ({AIGC}): A
  survey,'' \emph{arXiv preprint arXiv:2304.06632}, 2023.

\bibitem{janowicz2020geoai}
K.~Janowicz, S.~Gao, G.~McKenzie, Y.~Hu, and B.~Bhaduri, ``Geoai: spatially
  explicit artificial intelligence techniques for geographic knowledge
  discovery and beyond,'' pp. 625--636, 2020.

\bibitem{zhou2012geo}
Y.~Zhou and J.~Luo, ``Geo-location inference on news articles via multimodal
  plsa,'' in \emph{the 20th ACM International Conference on Multimedia}, 2012,
  pp. 741--744.

\bibitem{park2021review}
J.~Park and D.~W. Goldberg, ``A review of recent spatial accessibility studies
  that benefitted from advanced geospatial information: multimodal
  transportation and spatiotemporal disaggregation,'' \emph{ISPRS International
  Journal of Geo-information}, vol.~10, no.~8, p. 532, 2021.

\bibitem{zhu2018big}
L.~Zhu, F.~R. Yu, Y.~Wang, B.~Ning, and T.~Tang, ``Big data analytics in
  intelligent transportation systems: A survey,'' \emph{the Transactions on
  Intelligent Transportation Systems}, vol.~20, no.~1, pp. 383--398, 2018.

\bibitem{arfat2020big}
Y.~Arfat, S.~Usman, R.~Mehmood, and I.~Katib, ``Big data for smart
  infrastructure design: Opportunities and challenges,'' \emph{Smart
  Infrastructure and Applications: Foundations for Smarter Cities and
  Societies}, pp. 491--518, 2020.

\bibitem{joseph2013big}
R.~C. Joseph and N.~A. Johnson, ``Big data and transformational government,''
  \emph{It Professional}, vol.~15, no.~6, pp. 43--48, 2013.

\bibitem{akter2019big}
S.~Akter and S.~F. Wamba, ``Big data and disaster management: a systematic
  review and agenda for future research,'' \emph{Annals of Operations
  Research}, vol. 283, pp. 939--959, 2019.

\bibitem{rathore2016urban}
M.~M. Rathore, A.~Ahmad, A.~Paul, and S.~Rho, ``Urban planning and building
  smart cities based on the internet of things using big data analytics,''
  \emph{Computer Networks}, vol. 101, pp. 63--80, 2016.

\bibitem{runting2020opportunities}
R.~K. Runting, S.~Phinn, Z.~Xie, O.~Venter, and J.~E. Watson, ``Opportunities
  for big data in conservation and sustainability,'' \emph{Nature
  Communications}, vol.~11, no.~1, p. 2003, 2020.

\bibitem{jin2020big}
C.~Jin, Y.~Bouzembrak, J.~Zhou, Q.~Liang, L.~M. Van Den~Bulk, A.~Gavai, N.~Liu,
  L.~J. Van Den~Heuvel, W.~Hoenderdaal, and H.~J. Marvin, ``Big data in food
  safety-a review,'' \emph{Current Opinion in Food Science}, vol.~36, pp.
  24--32, 2020.

\bibitem{garrett2022climate}
K.~Garrett, D.~Bebber, B.~Etherton, K.~Gold, A.~Plex~Sul{\'a}, and M.~G.
  Selvaraj, ``Climate change effects on pathogen emergence: Artificial
  intelligence to translate big data for mitigation,'' \emph{Annual Review of
  Phytopathology}, vol.~60, pp. 357--378, 2022.

\bibitem{wiegand2007task}
N.~Wiegand and C.~Garc{\'\i}a, ``A task-based ontology approach to automate
  geospatial data retrieval,'' \emph{Transactions in GIS}, vol.~11, no.~3, pp.
  355--376, 2007.

\bibitem{huang2016memory}
W.~Huang, L.~Meng, D.~Zhang, and W.~Zhang, ``In-memory parallel processing of
  massive remotely sensed data using an apache spark on hadoop yarn model,''
  \emph{IEEE Journal of Selected Topics in Applied Earth Observations and
  Remote Sensing}, vol.~10, no.~1, pp. 3--19, 2016.

\bibitem{hoh2007preserving}
B.~Hoh, M.~Gruteser, H.~Xiong, and A.~Alrabady, ``Preserving privacy in gps
  traces via uncertainty-aware path cloaking,'' in \emph{the 14th ACM
  Conference on Computer and Communications Security}, 2007, pp. 161--171.

\bibitem{hiemenz2019dynamic}
B.~Hiemenz and M.~Kr{\"a}mer, ``Dynamic searchable symmetric encryption for
  storing geospatial data in the cloud,'' \emph{International Journal of
  Information Security}, vol.~18, pp. 333--354, 2019.

\bibitem{budgaga2017framework}
W.~Budgaga, M.~Malensek, S.~Lee~Pallickara, and S.~Pallickara, ``A framework
  for scalable real-time anomaly detection over voluminous, geospatial data
  streams,'' \emph{Concurrency and Computation: Practice and Experience},
  vol.~29, no.~12, p. e4106, 2017.

\bibitem{bereta2016ontop}
K.~Bereta and M.~Koubarakis, ``Ontop of geospatial databases,'' in \emph{the
  15th International Semantic Web Conference}.\hskip 1em plus 0.5em minus
  0.4em\relax Springer, 2016, pp. 37--52.

\bibitem{amiri2021sharper}
M.~J. Amiri, D.~Agrawal, and A.~El~Abbadi, ``Sharper: Sharding permissioned
  blockchains over network clusters,'' in \emph{the International Conference on
  Management of Data}, 2021, pp. 76--88.

\bibitem{christoforaki2011text}
M.~Christoforaki, J.~He, C.~Dimopoulos, A.~Markowetz, and T.~Suel, ``Text vs.
  space: efficient geo-search query processing,'' in \emph{the 20th
  International Conference on Information and Knowledge Management}, 2011, pp.
  423--432.

\bibitem{dayarathna2018recent}
M.~Dayarathna and S.~Perera, ``Recent advancements in event processing,''
  \emph{ACM Computing Surveys}, vol.~51, no.~2, pp. 1--36, 2018.

\bibitem{breunig2020geospatial}
M.~Breunig, P.~E. Bradley, M.~Jahn, P.~Kuper, N.~Mazroob, N.~R{\"o}sch,
  M.~Al-Doori, E.~Stefanakis, and M.~Jadidi, ``Geospatial data management
  research: Progress and future directions,'' \emph{International Journal of
  Geo-Information}, vol.~9, no.~2, p.~95, 2020.

\bibitem{zhang2013semantic}
Y.~Zhang, Y.-Y. Chiang, P.~Szekely, and C.~A. Knoblock, ``A semantic approach
  to retrieving, linking, and integrating heterogeneous geospatial data,'' in
  \emph{Joint Proceedings of the Workshop on AI Problems and Approaches for
  Intelligent Environments and Workshop on Semantic Cities}, 2013, pp. 31--37.

\bibitem{wang2019integrated}
S.~Wang, Y.~Zhong, and E.~Wang, ``An integrated gis platform architecture for
  spatiotemporal big data,'' \emph{Future Generation Computer Systems},
  vol.~94, pp. 160--172, 2019.

\bibitem{karalis2019extending}
N.~Karalis, G.~Mandilaras, and M.~Koubarakis, ``Extending the yago2 knowledge
  graph with precise geospatial knowledge,'' in \emph{the 18th International
  Semantic Web Conference}.\hskip 1em plus 0.5em minus 0.4em\relax Springer,
  2019, pp. 181--197.

\bibitem{liu2021heterogeneous}
J.~Liu, H.~Liu, X.~Chen, X.~Guo, Q.~Zhao, J.~Li, L.~Kang, and J.~Liu, ``A
  heterogeneous geospatial data retrieval method using knowledge graph,''
  \emph{Sustainability}, vol.~13, no.~4, p. 2005, 2021.

\bibitem{dsouza2021worldkg}
A.~Dsouza, N.~Tempelmeier, R.~Yu, S.~Gottschalk, and E.~Demidova, ``{WorldKG}:
  A world-scale geographic knowledge graph,'' in \emph{the 30th ACM
  International Conference on Information \& Knowledge Management}, 2021, pp.
  4475--4484.

\bibitem{moser2010beyond}
J.~Moser, F.~Albrecht, and B.~Kosar, ``Beyond visualisation--{3D} {GIS}
  analyses for virtual city models,'' \emph{International Archives of the
  Photogrammetry, Remote Sensing and Spatial Information Sciences}, vol.~38,
  no.~4, p. W15, 2010.

\bibitem{sun2022metaverse}
J.~Sun, W.~Gan, H.-C. Chao, and P.~S. Yu, ``Metaverse: Survey, applications,
  security, and opportunities,'' \emph{arXiv preprint arXiv:2210.07990}, 2022.

\bibitem{jiang2015geospatial}
B.~Jiang, ``Geospatial analysis requires a different way of thinking: The
  problem of spatial heterogeneity,'' \emph{GeoJournal}, vol.~80, pp. 1--13,
  2015.

\bibitem{mai2023csp}
G.~Mai, N.~Lao, Y.~He, J.~Song, and S.~Ermon, ``{CSP}: Self-supervised
  contrastive spatial pre-training for geospatial-visual representations,'' in
  \emph{International Conference on Machine Learning}.\hskip 1em plus 0.5em
  minus 0.4em\relax PMLR, 2023, pp. 23\,498--23\,515.

\bibitem{yang2022sparse}
F.~Yang and C.~Ma, ``Sparse and complete latent organization for geospatial
  semantic segmentation,'' in \emph{the IEEE/CVF Conference on Computer Vision
  and Pattern Recognition}, 2022, pp. 1809--1818.

\bibitem{tam2021adaptive}
P.~Tam, S.~Math, C.~Nam, and S.~Kim, ``Adaptive resource optimized edge
  federated learning in real-time image sensing classifications,'' \emph{IEEE
  Journal of Selected Topics in Applied Earth Observations and Remote Sensing},
  vol.~14, pp. 10\,929--10\,940, 2021.

\bibitem{sprague2018asynchronous}
M.~R. Sprague, A.~Jalalirad, M.~Scavuzzo, C.~Capota, M.~Neun, L.~Do, and
  M.~Kopp, ``Asynchronous federated learning for geospatial applications,'' in
  \emph{Joint European Conference on Machine Learning and Knowledge Discovery
  in Databases}.\hskip 1em plus 0.5em minus 0.4em\relax Springer, 2018, pp.
  21--28.

\bibitem{algarni2023p3s}
F.~Algarni, M.~A. Khan, W.~Alawad, and N.~B. Halima, ``{P3S}: Pertinent privacy
  preserving scheme for remotely sensed environmental data in smart cities,''
  \emph{IEEE Journal of Selected Topics in Applied Earth Observations and
  Remote Sensing}, pp. 5905--5918, 2023.

\end{thebibliography}

\end{document}